\documentclass[3p,fleqn]{elsarticle}

\usepackage{lineno}
\usepackage{amssymb}
\usepackage{amsmath,amsthm,mathtools}

\usepackage{moreverb}
\usepackage{multirow}
\usepackage{subfig}
\usepackage{graphicx}
\usepackage{epstopdf}

\usepackage{algorithm}
\usepackage{algorithmic}

\usepackage{alltt}

\modulolinenumbers[5]

\usepackage{color} 
\usepackage{listings}

\journal{Elsevier}

\begin{document}

\begin{frontmatter}

\title{An Approach for Accelerating Incompressible Turbulent Flow Simulations
Based on Simultaneous Modelling of Multiple Ensembles}

\author{Boris~I.~Krasnopolsky}
\ead{krasnopolsky@imec.msu.ru}

\address{Laboratory of General Aerodynamics, Institute of Mechanics,
Lomonosov Moscow State University, 119192, Moscow, Michurinsky ave.~1, Russia}

\begin{abstract}
The present paper deals with the problem of improving the efficiency of large
scale turbulent flow simulations. \textcolor{black}{The high-fidelity methods for
modelling turbulent flows become available for a wider range of applications
thanks to the constant growth of the supercomputers performance, however, they
are still unattainable for lots of real-life problems.} The key shortcoming of
these methods is related to the need of simulating a long time integration
interval to collect reliable statistics, while the time integration process is
inherently sequential.

The novel approach with modelling of multiple flow states is discussed in the
paper. The suggested numerical procedure allows to
\textcolor{black}{parallelize} the integration in time by the cost of
additional computations. Multiple realizations of the same turbulent flow are
performed simultaneously. This allows to use more efficient implementations of
numerical methods for solving systems of linear algebraic equations with
multiple right-hand sides, operating with blocks of vectors. The simple
theoretical estimate for the expected simulation speedup, accounting the penalty
of additional computations and the linear solver performance improvement, is
presented. \textcolor{black}{The two problems of modelling turbulent flows in a plain
channel and in a channel with a matrix of wall-mounted cubes are used to
demonstrate the correctness of the proposed estimates and efficiency of the
suggested approach as a whole.} The simulation speedup by a factor of~2 is shown.
\end{abstract}


\end{frontmatter}


\section{Introduction}

The modelling of turbulent flows is one of the typical applications for high
performance computing (HPC) systems. \textcolor{black}{The accurate
eddy-resolving simulations of turbulent flows are characterized by huge
computational grids and long time integration intervals, making them extremely
time-consuming to solve. Despite the constant growth of the computational power
of HPC systems, the use of high-fidelity methods} is still unattainable for lots
of real-life applications. This motivates the researchers to develop new
computational algorithms and adapt the known algorithms to the modern HPC
systems.

The high-order time integration schemes are typically used to increase the
accuracy of turbulent flow simulations with eddy-resolving methods.
\textcolor{black}{The 3-th or 4-th order explicit or semi-implicit Runge-Kutta
schemes} (e.g.,~\cite{ref:Spalart1991, ref:Nikitin2006, ref:Trias2011}) or 2-nd
order Adams-Bashforth/Crank-Nicolson schemes~\cite{ref:Moin1978, ref:Moin1982,
ref:Mahesh2004, ref:Kim2000} are among the widely used ones for time integration
of incompressible flows. In these schemes, the preliminary velocity
distributions are obtained from the Navier-Stokes equations, and the continuity
equation, transformed to the elliptic pressure Poisson equation, is used
\textcolor{black}{to enforce a divergence-free velocity field}. Commonly, the
solution of elliptic equations and corresponding systems of linear algebraic
equations (SLAEs) is a complicated and challenging problem.
The time to compute this stage can take up to 95\% of the overall simulation
time.

For a limited number of problems with regular computational domains, solved on
structured grids, the direct methods for solving SLAEs can be used
(e.g.,~\cite{ref:Swarztrauber1974, ref:Gorobets2010}). Otherwise, the iterative
methods are the good candidates to solve SLAEs with matrices of the general form.
The multigrid methods~\cite{ref:Trottenberg2001} or Krylov subspace methods
(e.g., BiCGStab~\cite{ref:Vorst1992, ref:Yang2002, ref:Krasnopolsky2010},
GMRES~\cite{ref:Saad1986}) with multigrid preconditioners are the popular ones
to solve the corresponding systems. The advantages of these methods are
related to their robustness and excellent scalability
potential~\cite{ref:Baker2012}. Mathematically, these methods consist of a
combination of linear operations with dense vectors, ${\bf z} = a{\bf x} + b{\bf
y}$, scalar products, $a = ({\bf x},{\bf y})$, and sparse matrix-dense vector
multiplications (SpMV), ${\bf y} = {\bf A x}$. While the linear operations with
dense vectors and scalar products are easily vectorized by compilers and
acceptable performance is achieved, the performance of SpMV operations is
dramatically lower. The sparse matrix-vector multiplication is a memory-bound
operation with extremely low arithmetic intensity. The real performance of
linear algebra algorithms with sparse matrices of the general form does not exceed
several percent of the peak performance~\cite{ref:Gropp1999, ref:Williams2007,
ref:Williams2009}. The optimizations of operations with sparse matrices, which
are related to both optimization of matrix storage formats and implementation
aspects are a topic of continuous research for many years
(e.g.,~\cite{ref:Buluc2009, ref:Yzelman2009, ref:Martone2014,
ref:Kreutzer2014}).

The performance of SpMV-like operations can be significantly improved if
applied to a block of dense vectors simultaneously (generalized SpMV, GSpMV),
${\bf Y} = {\bf A X}$, where $\bf X$ and $\bf Y$ are dense
matrices~\cite{ref:Gropp1999, ref:Liu2012, ref:Aktulga2014, ref:Imamura2016}.
Generally, the performance gain of GSpMV operation with $m$ vectors compared to
$m$ successive SpMV operations is achieved due to two main factors: (i)
the reduction of the memory traffic to load the matrix $\bf{A}$ from the memory
(the matrix is read only once) and (ii) vectorization improvement for GSpMV
operation.

Despite the significant performance advantage of matrix-vector operations with
blocks of vectors over the single-vector operations, the GSpMV-like operations
are rarely used in real computations. This fact is a consequence of the
numerical algorithms design: most of them operate with single vectors only.
Among the several exceptions allowing to exploit the GSpMV-like operations are
the applications with natural parallelism over the right-hand sides (RHS) when
solving SLAEs. For example, in structural analysis applications, the solution of
SLAE with multiple RHS arise for multiple load vectors~\cite{ref:Feng1995}. In
computational fluid dynamics, the operations with groups of vectors can be used
for solving Navier-Stokes equations (e.g., for computational algorithms
operating with collocated grids and explicit discretization of nonlinear terms).
For multiphase flows, the concentration transport equations for the phases
forming the carrier fluid can also be solved in a single run.

In addition, several articles are focused on the attempts to modify the
computational procedure in order to organize the computations with groups of
vectors. For example, the modified Stokesian dynamics method for the simulation
of the motion of macromolecules in the cell, exploiting the advantages of
operations with groups of vectors, is presented in~\cite{ref:Liu2012}.
\textcolor{black}{The benefits of ensemble computing for current HPC systems are
outlined in~\cite{ref:Imamura2016}. The authors suggested to perform together
several incompressible flow simulations (e.g. applications with varying initial
or boundary conditions), which have a common sparse matrix derived from the
pressure Poisson equation. This modification allows to combine multiple
solutions of the pressure Poisson equation in a single operation with multiple
right-hand sides. The 2.4 and 7.6~times speedup for the successive
over-relaxation method used to solve the corresponding SLAEs with up to 128~RHS
vectors on Intel and Sparc processors are reported by the authors.
}

\textcolor{black}{The problem of long time integration for high-fidelity
turbulent flow simulations is discussed in~\cite{ref:Makarashvili2017}. The
authors suggested to combine the conventional time averaging approach for the
statistically steady turbulent flows with the ensemble averaging. Several
turbulent flow realizations are performed independently with a shorter time
integration interval, and the obtained results are averaged at the end of the
simulation. Scheduling additional resources to perform each of the flow
realizations, the proposed approach allows to speedup the simulations beyond the
strong scaling limit by the extra computational costs.
}

\textcolor{black}{The current paper discusses an idea of combining the ensemble
averaging for statistically steady turbulent flow
simulations~\cite{ref:Makarashvili2017} with the simultaneous modelling of
multiple flow realizations~\cite{ref:Imamura2016} to speedup the high-fidelity
simulations. The focus is on the reduction of the overall
computational costs for the corresponding simulations.} The paper provides the
modified computational procedure to model incompressible turbulent flows,
allowing to utilize the operations with groups of vectors. For the sake of
simplicity, the further narration is focused on the direct numerical
simulation (DNS) aspects; however, the proposed methodology can be applied ``as
is'' for the large eddy simulation (LES) computations.

The paper is organized as follows. The motivating observations and theoretical
prerequisites of the proposed computational procedure are stated in the second
section. The third section contains description of the numerical methods and
computational codes used to simulate turbulent flows. Numerical results
validating the theoretical estimates and demonstrating the advantages of the
proposed algorithm are presented in the fourth section.
\textcolor{black}{The fifth section discusses the applicability of the
proposed approach to other mathematical models and techniques that can be used
for further efficiency improvements.}


\section{Preliminary observations and theoretical estimates}
\label{sec:estimates}

\subsection{Time-averaged and ensemble-averaged statistics}

\begin{figure}[t]
  \centering
    \includegraphics[width=10cm]{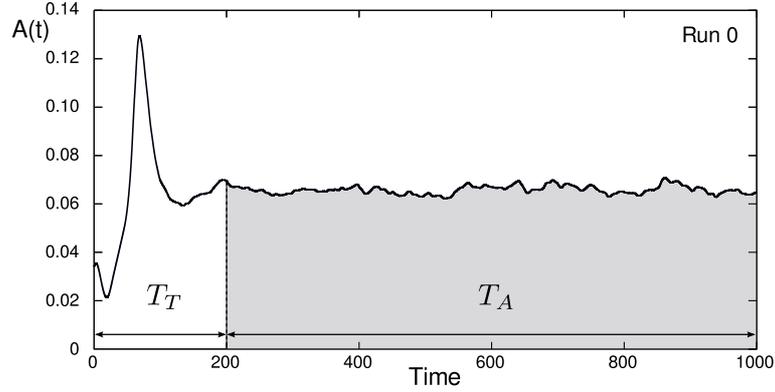}
  \caption{Integral velocity perturbations amplitude~\cite{ref:Voronova2006} for
  DNS of turbulent flow in a straight pipe, $\mbox{Re}_D=6000$; averaging over
  the single flow.}
  \label{fig:vel_profile1}
\end{figure}

\textcolor{black}{The DNS/LES of turbulent flow comprises integration in
time. For the statistically steady turbulent flow the time integration process
typically consists of two stages (Figure~\ref{fig:vel_profile1}).
The first stage, $T_T$, reflects transition from the initial state to the
statistically steady regime. The second stage, $T_A$, includes the averaging of
instantaneous velocity fields and collecting the turbulent statistics. The ratio
of these intervals may vary significantly depending on the specific application.
For example, following the ergodicity hypothesis~\cite{ref:Galanti2004,
ref:Tsinober2009}, the flows with homogeneous directions can be averaged along
these directions thus significantly reducing the time averaging interval.
Opposite, the applications with complex geometries often need to perform much
larger time averaging intervals to obtain reliable statistics compared to the
time to reach the statistical equilibrium.
}

\begin{figure}[t]
  \centering
    \includegraphics[width=10cm]{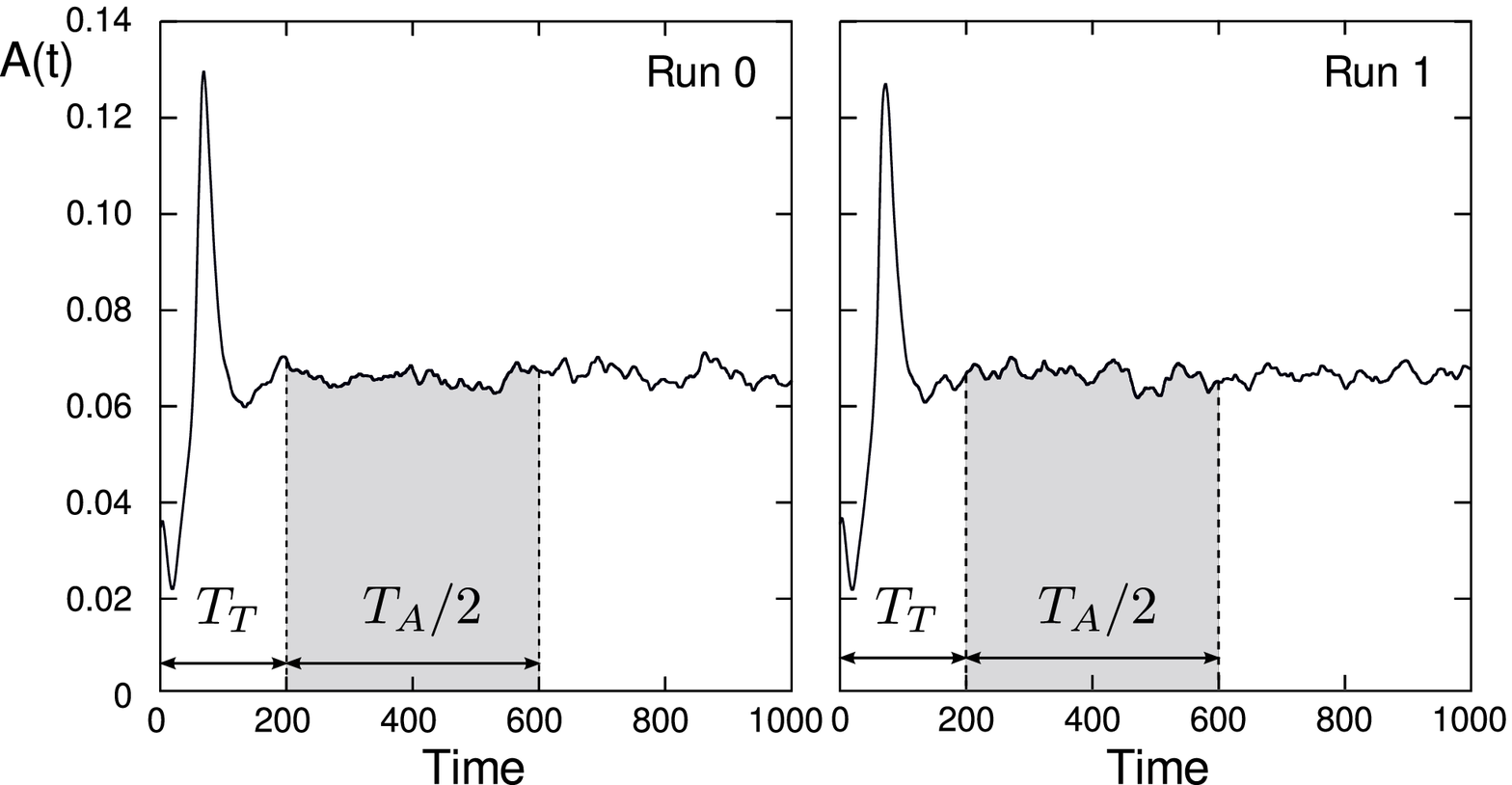}
  \caption{Integral velocity perturbations amplitude~\cite{ref:Voronova2006} for
  DNS of turbulent flow in a straight pipe, $\mbox{Re}_D=6000$; averaging over
  two flows.}
  \label{fig:vel_profile2}
\end{figure}

\textcolor{black}{The time averaging to collect turbulent statistics is applied
to statistically steady turbulent flows. In case the flow considered is not
statistically steady, the ensemble averaging must be performed. The time
averaging of statistically steady flows can be combined with the ensemble
averaging~\cite{ref:Makarashvili2017}. Instead of averaging in time over the
single flow with time interval $T_A$, $m$ flows can be averaged over the
interval $T_A/m$ (Figure~\ref{fig:vel_profile2}). The flow realizations are
integrated over the time interval $T_m = T_T + T_A/m$. The parallelization in
time increases the overall integration interval, which becomes equal to
$T_{total} = m \cdot T_m = m T_T + T_A$. The additional $(m-1)$ initial flow
transformations leading to the extra time integration interval $(m-1) T_T$
provide a noticeable drawback, generally making the advantages of simultaneous
modelling of multiple flow realizations a~priori not evident.
}

\textcolor{black}{ The ensemble averaging provides the requirement for the
initial turbulent flow states to be uncorrelated. This, however, can be
satisfied by choosing a proper transition interval. The divergence of two
slightly different turbulent flows is exponential and the corresponding
divergence growth rate in the phase-state is determined by the Lyapunov exponent
(see, e.g.~\cite{ref:Nikitin2009, ref:Nastac2017}). Thus, increasing the length
of the transition interval one can obtain uncorrelated initial turbulent flow
states.}

The speedup for simulation with averaging over multiple flows can easily be
achieved by increasing the number of scheduled computational resources. Each
flow state can be performed independently with only final post-processing of
results~\cite{ref:Makarashvili2017}. In practice, the significant increase in
computational resources for the simulation may be impossible for technical or
economic reasons. The current paper focuses on attempt to speedup the
simulations by increasing the efficiency of computations on the same hardware
resources, discussing an idea of simultaneous modelling of multiple independent
flow realizations.

\subsection {Time integration schemes}

\textcolor{black}{The system of governing equations for direct numerical
simulation of isothermal incompressible viscous flows consists of Navier-Stokes
and continuity equations:
\begin{gather}
\frac{\partial {\bf{u}}}{\partial t} + {\bf{u}} \cdot \nabla {\bf{u}} =
 - \nabla p + \frac{1}{\mbox{Re}} \nabla^2 {\bf{u}}, \label{eq:NS} \\
\nabla \cdot {\bf{u}} = 0, \label{eq:cont}
\end{gather}
where $\bf{u}$ is the velocity field, $p$ is the pressure, and $t$ is the time.
Equations \eqref{eq:NS}-\eqref{eq:cont} are written in the dimensionless form.
The nondimensional Reynolds number is defined as $\mbox{Re} = U L / \nu$, where
$U$ is the characteristic velocity, $L$ is the characteristic length scale, and
$\nu$ is the kinematic viscosity of the fluid.}

Let us consider the time integration schemes used for direct numerical
simulations. The high order explicit or semi-implicit Runge-Kutta schemes
(e.g.,~\cite{ref:Spalart1991, ref:Nikitin2006, ref:Trias2011}) are typically
used for high-fidelity turbulent flow simulations in order to reduce the
numerical dissipation effects. \textcolor{black}{These schemes consist of
several substeps (the number of substeps depends on the approximation order of
the numerical scheme). With a certain degree of simplification, each substep is
comprised of predicting preliminary velocity field from~\eqref{eq:NS}, and
projecting obtained velocity field on the divergence-free space by
solving~\eqref{eq:cont}, transformed to the pressure Poisson equation:
\begin{gather}
\tilde{\bf u} = {\bf u}^{*} -\tau \nabla \tilde p, \label{eq:eq2}\\
\nabla^2 \tilde p = \frac{1}{\tau} \, \nabla \cdot {\bf{u}}^*.\label{eq:eq3}
\end{gather}
Here ${\bf u}^{*}$ is the preliminary velocity, $\tilde{\bf u}$ is the
divergence-free velocity at the end of the substep, $\tilde{p}$ is the pressure at
the end of the substep, and $\tau$ is a variable related to the integration
step.}

The most time-consuming part for obtaining solution
of~\eqref{eq:NS}-\eqref{eq:cont} at the next time step relates to the solution
of pressure Poisson equation~\eqref{eq:eq3}. One can see the spatial
discretization and the corresponding SLAE matrix for the elliptic equation
remain constant \textcolor{black}{in the case of the grid remaining unchanged}
during the simulation. The matrix coefficients are also independent of the
velocity distribution and the divergence of the preliminary velocity field arise
on the right-hand side of the SLAE only. This means having $m$ different
preliminary velocity fields in~\eqref{eq:eq2}, $m$ pressure fields can be
performed simultaneously by solving the SLAE with $m$ RHS vectors.

Generally speaking, the suggested modification of the computational procedure
can also be extended on the case of moving grids, when the grid transformation
is a function of the time only (e.g., the grids with moving regions), but not
the instantaneous velocity fields (e.g., adaptively refined meshes). This,
however, provides an additional limitation to the time integration: all the
flows must be synchronized by the time step and in time. This limitation may
reduce the expected gains for the proposed transformation of the computational
algorithm and should be analyzed in each case.

\subsection {Theoretical estimates}
\label{sec:theoretical_estimates}

The proposed above modification of the numerical procedure for simultaneous
modelling of multiple turbulent flow states allows to obtain higher performance
for solving SLAEs of the pressure Poisson equation (which are typically
dominating in the overall computing time), but needs some additional
computations. Let us consider the basic theoretical estimates for the proposed
modification to analyze possible gains. The value of interest is the parameter
$P_m$, the overall simulation speedup for the simultaneous modelling of $m$
turbulent flow states compared to the simulation of the single turbulent flow
state.
By this definition, the parameter is a ratio of the time to compute the whole
simulation with single flow state, $\mathcal {T}_1$, to the one with $m$ flow
states, $\mathcal {T}_m$,
\begin{equation}
P_m = \frac {\mathcal {T}_1}{\mathcal {T}_m}.\label{eq:est}
\end{equation}

\textcolor{black}{The Runge-Kutta time integration schemes have a fixed number
of substeps thus providing close to constant computational complexity per time
step during the simulation.} Then, the simulation time can be estimated as:
\begin{equation}
\mathcal{T}_m = N_m t_m,
\end{equation}
where $N_m$ is the number of time steps to be modelled, and $t_m$~is the
computation time per integration step with $m$~flow states. The time integration
in DNS/LES is typically performed with constant or predominantly constant time
step, $\tau$. As a result, the number of time steps can be expressed through the
physical simulation time, $T_m$:
\begin{equation}
N_m = \frac{T_m}{\tau},
\end{equation}
where $T_m = T_T + T_A / m$. Introducing the parameter $\beta = T_A / T_T$, the
ratio of time averaging to transition interval, the expression~\eqref{eq:est}
can be represented as:
\begin{equation}
P_m = \frac{1+\beta}{m+\beta} \frac{m t_1}{t_m}. \label{eq:est2}
\end{equation}
The first multiplier in~\eqref{eq:est2} reflects the overhead to perform
additional $(m-1)$~flow states transitions and the second multiplier accounts
the simulation speedup for simultaneous modelling of $m$~flow states compared to
$m$ successive computations.

The computation time per integration step is approximated as a sum of two
factors: the time to solve SLAEs for the pressure Poisson equation, $t^S$, and
the time to construct the temporal and spatial discretization of the governing
equations, $t^D$:
\begin{equation}
t_m = t_m^S + t_m^D.
\end{equation}
The time to construct the discretizations is expressed in proportion to the
number of flow states,
\begin{equation}
t_m^D = m t_1^D,
\end{equation}
as no gain is expected here due to the simulations with multiple flow states.
Introducing the parameter $\theta$, the fraction of the SLAE solver time in the
integration step execution time, $\theta = t_1^S / t_1$, one can obtain the
following expression:
\begin{equation}
\frac{m t_1}{t_m} = \frac{1}{\theta \left( \frac{m t_1^S}{t_m^S} \right)^{-1} +
\left( 1-\theta \right)}.\label{eq:est2_1}
\end{equation}

To complete, the expression for the SLAE solver time as a function of the number
of RHS vectors must be provided. The iterative methods for solving elliptic
SLAEs with matrix of the general form, specifically Krylov subspace and
multigrid methods, are based on SpMV operations and operations with dense
vectors, and the former ones typically dominate in the solution time. With some
assumptions, the time to solve the SLAE with multiple RHS vectors can be
approximated in proportion to the time for GSpMV operation with $m$~vectors. The
GSpMV is a memory-bound operation and its execution time coincides with the
volume of data transfers with the memory. The total volume of data transfers
depends on the matrix storage format, and the choice of the optimal one is
affected by lots of factors, i.e. the matrix size, the number of nonzero
elements, subblock structure of nonzero elements, matrix topology, etc.

For the CRS data format~\cite{ref:Saad2003}, which, despite of its simplicity,
is still among the most popular ones in real applications, the volume of data
transfers can be calculated explicitly. Let us consider the square
matrix~\textbf{A} with $n$~rows and $nnz$~nonzero elements, and two dense
matrices \textbf{X} and \textbf{Y} with $n$~rows and $m$~columns. The matrix
\textbf{A} is stored in CRS format and the dense matrices \textbf{X} and
\textbf{Y} are stored row-wise in single vectors. For the sake of simplicity
the integer numbers are assumed 4~bytes and floating point numbers are 8~bytes.
The C-like code for the GSpMV operation $\bf{Y} = \bf{A} \bf{X}$ with $m$~RHS
vectors is presented in Figure~\ref{fig:GSpMV_algorithm}.

\begin{figure}[htb]
\begin{lstlisting}
GSpMV (Matrix *M, double *X, double *Y, int m)
{
 int i, j1, j2, k, l;
 double sum[m];
 
 j1 = M->row[0];                     // (1)
 for(i = 0; i < M->n; i++)
 {
  j2 = M->row[i+1];                  // (1)
  for(l = 0; l < m; l++)
   sum[l] = 0.;

  for(j = j1; j < j2; j++) 
  {
   k = M->col[j];                    // (2)
   for(l = 0; l < m; l++)
    sum[l] += M->val[j] * X[k*m+l];  // (3) & (4)
  }
  
  j1 = j2;
  for(l = 0; l < m; l++)
   Y[i*m+l] = sum[l];                // (5)
 }
}
\end{lstlisting}
\caption{C-like code for GSpMV operation with matrix stored in CRS format.}
\label{fig:GSpMV_algorithm}
\end{figure}

The memory traffic produced by the GSpMV operation consists of read~/~write
operations with 5~arrays: (1)~$(n+1)$ integers reads from the array
\textit{row}; (2)~$nnz$ integers reads from the array \textit{col}; (3)~$nnz$
floating point numbers reads from the array \textit{val}; (4)~$m \cdot nnz$
floating point numbers reads from the array \textit{X}; (5)~$m \cdot n$ floating
point numbers writes to the array \textit{Y}. In total, the GSpMV operation with
$m$ vectors produces the data transfer with the memory equal to (single memory
read in~(1) is omitted):
\begin{equation}
{\sum}_{m} = 4n(2m+1) + 12nnz + 8 m \cdot nnz.
\end{equation}
The corresponding memory transfer gain is:
\begin{equation}
\frac{m {\sum}_{1}}{{\sum}_{m}} = \frac{m \left(5C + 3\right)}{2m
\left(C+1\right) +3C +1}, \label{eq:mem_GSpMV_full}
\end{equation}
where $C = nnz / n$ is the average number of nonzero elements per matrix row. 
Taking into account, that typically $C \gg 1$, \eqref{eq:mem_GSpMV_full} can be
reduced to a trivial expression:
\begin{equation}
\frac{m {\sum}_{1}}{{\sum}_{m}} \approx \frac{5 m}{2m
+ 3}. \label{eq:mem_GSpMV_compact}
\end{equation}
Using~\eqref{eq:mem_GSpMV_compact} as an estimate for the SLAE solver execution
times ratio in~\eqref{eq:est2_1}, the final expression can be achieved:
\begin{equation}
P_m = \frac{1+\beta}{m+\beta} \frac{5 m}{5m - 3\theta \left( m-1 \right)}. \label{eq:est3}
\end{equation}
The presented estimate~\eqref{eq:est3} is a function of three parameters:
the times ratio $\beta$, the fraction of the SLAE solver time in the computation
time per integration step $\theta$, and the number of averaging ensembles $m$.
The parameter $\beta$ is a characteristic of the specific application modelled
and is an input parameter for this estimate. The parameter $\theta$ is a
characteristic of the numerical schemes and methods used for spatial and
temporal discretization and solving SLAEs, and is also an input parameter.
The parameter $m$ is a free parameter and it is reasonable to be chosen in order
to maximize the simulation speedup. \textcolor{black}{Differentiating
expression~\eqref{eq:est3} over $m$ one can obtain:
\begin{equation}
m^* = \sqrt{\frac{3 \beta \theta}{5 - 3 \theta}},
\end{equation}
and the nearest integer value to $m^*$ can be treated as an expected
optimal number of flow states, maximizing the overall simulation speedup.}

The two limiting cases for~\eqref{eq:est3} can be considered. In case the time
averaging dominates over the transition, $\beta \gg 1$, \eqref{eq:est3}~reduces
to
\begin{equation}
P_m = \frac{5 m}{5m - 3\theta \left( m-1 \right)},
\end{equation}
that suggests more than 2-fold speedup can be achieved by the simultaneous
modelling of several flow states. In the opposite case when the transition
dominates over the time averaging, $\beta \ll 1$,~\eqref{eq:est3} transforms to
\begin{equation}
P_m = \frac{5}{5m - 3\theta \left( m-1 \right)}.\label{eq:slowdown}
\end{equation}
Expression~\eqref{eq:slowdown} means that the overall simulation slowdown is
expected and the proposed idea is inapplicable for this type of problems.

\begin{figure}[htbp]
  \centering
  \includegraphics[width=10cm]{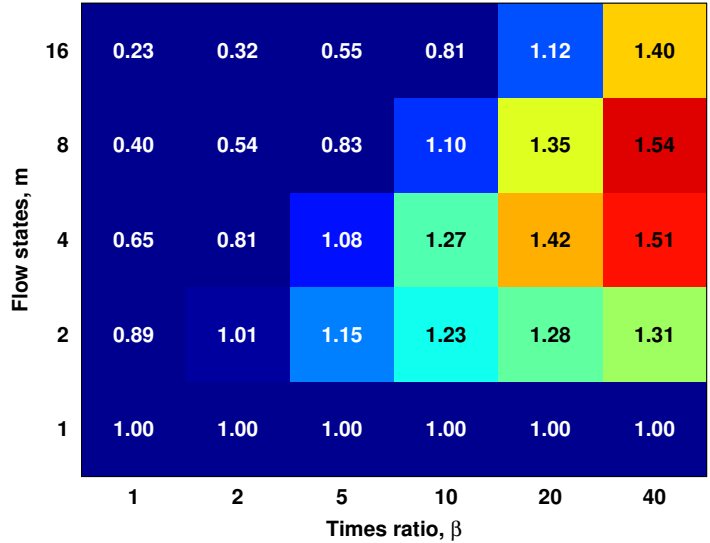}
  \caption{The estimated simulation speedup, $P_m$, as a function of the times
  ratio, $\beta$, and the number of modelled flow states, $m$.}
  \label{fig:P_estimate}
\end{figure}

Several reference values for the estimated simulation speedup with modelling
multiple flow states are presented in Figure~\ref{fig:P_estimate}. The parameter
$\theta$ here is set to $\theta = 0.85$, which is an average of the values
obtained in the numerical experiments \textcolor{black}{considered in the
validation section}, where the SLAE solver input to the overall simulation time
varied from 75 to~95\%. The second parameter $\beta$ varies from 1 to~40. The
presented figure demonstrates that the optimal number of flow states depends on
the times ratio $\beta$ and changes from $m=2$ for $\beta = 2 - 5$ to $m=4$ for
$\beta = 10 - 20$ and to $m=8$ for $\beta = 40$. The expected speedup for $\beta
= 5 - 40$ is about 15-54\%.

It should be noted that the proposed estimate for the SLAE solver speedup due
to its simplicity ignores several important aspects, like vectorization,
regularization of memory access pattern and improvements of cache miss rate.
Solution of SLAE with multiple RHS vectors also has a significant advantage for
parallel computations, allowing to increase the amount of computations used to hide the
MPI communications latency for non-blocking point-to-point communications. The
number of RHS vectors affects the message sizes, but not the total amount of
the messages, thus only slightly increasing the message transfer time.

\textcolor{black}{The physical constraint of choosing the number of flow states
should be considered in addition to the estimate~\eqref{eq:est3}. The time
averaging interval per each flow state, $T_A/m$, must be preserved larger than
the characteristic time scale of the modelled turbulent flow.
This treatment provides the upper bound for the parameter~$m$. In general, the
typical time averaging interval $T_A$ is chosen to cover several tens to
hundreds of characteristic time scales in order to obtain reliable statistics.
The expected optimal number of flow states maximizing the simulation speedup is
typically of order $\mathcal{O}(1)$ (see~Figure~\ref{fig:P_estimate}). This
means the time averaging interval $T_A/m$ would generally cover at least several
characteristic time scales.}

\textcolor{black}{The use of computations with multiple RHS vectors increases the
memory consumption per each compute node. The memory consumption scales
proportionally to the number of flow states, except the information related to
the simulated problem statement, which can be stored only once.
This aspect may limit the applicability of the suggested approach in the case of
huge computational grids and low memory capacity of the compute nodes scheduled
for the simulation. However, the use of large grid blocks per node would require
extremely long execution times for the whole simulation. The codes for DNS/LES
usually demonstrate good scalability across multiple nodes with the granularity
of lower than 100~thousand cells per node (e.g.~\cite{ref:Borrell2016,
ref:Offermans2017}). This allows to store in the memory without in due
difficulties the data for several tens of flow states.}

Accordingly, the proposed simple estimates demonstrate the possibility of the
DNS/LES computations speedup by the simultaneous modelling of several
independent flow states with subsequent post-processing of simulation results.
While the speedup of only 15-54\% is predicted by the estimates, even these
results can be of high practical importance as typical high-fidelity turbulent
flow simulations for the objects with complex geometry \textcolor{black}{may
take months to complete}.


\section{Numerical methods and software}
\label{sec:num_methods}

To validate the proposed computational procedure and demonstrate the simulation
speedup with modelling of multiple flow states the corresponding application for
direct numerical simulation of turbulent flows and the SLAE solver were
developed. The DNS application is an extension of ``in-house'' code for
modelling incompressible turbulent flows. The code is based on the finite
difference scheme for structured grids, and operates with arbitrary curvilinear
orthogonal coordinates~\cite{ref:Nikitin2006a}. The second order central
difference scheme on the staggered grid, preserving the discrete kinetic energy
conservation property, is used for the spatial discretization. The time
integration is performed by 3-1/3 step semi-implicit Runge-Kutta scheme with
optimal time stepping algorithm~\cite{ref:Nikitin2006}.
\textcolor{black}{The message passing programming model (MPI) is used to
parallelize the computations. The simple 3D geometrical decomposition into
equally sized subdomains is applied to distribute the problem across the
computational processes.}

The SparseLinSol library~\cite{ref:Krasnopolsky2016}, containing a set of Krylov
subspace and multigrid iterative methods, is used to solve the systems of linear
algebraic equations. The corresponding extension for solving systems with
multiple RHS vectors was developed. The library is based on hybrid multilevel
parallel algorithms, accounting memory hierarchy optimal access pattern with
four logical levels: ``compute node / socket / numa-node / core'' for CPUs
with intra-node communications and synchronizations via POSIX Shared Memory.
\textcolor{black}{The library functionality allows efficient coupling with
applications designed under several parallel programming models, including the
pure MPI approach.}


\section{Validation results}
\label{sec:num_results}

\begin{table*}
\caption{Parameters of the computational grids.\label{tab:grid}}
\centering
\begin{tabular}{ | c | c | c | c |}
\hline
	             & Grid 1 & Grid 2 & Grid 3 \\
\hline
 	Overall & $144 \times 112 \times 144$ & $240 \times 168 \times 240$ & $360
 	\times 252 \times 360$ \\
\hline
	Cube    & $52 \times 48 \times 52$ & $100 \times 74 \times 100$ & $150 \times
	120 \times 150$ \\
\hline
	$\Delta x$, $\Delta z$   & $0.0065h - 0.054h$ & $0.0031h - 0.038h$ & $0.0024h -
	0.023h$
	\\
	$\Delta y$   & $0.0054h - 0.07h$  & $0.003h - 0.044h$ & $0.0023h - 0.03h$ \\
\hline
	$\Delta y^+_{top}$   & $1.50$  & $1.10$ & $0.98$ \\
\hline
\end{tabular}
\end{table*}

\begin{figure}[htb]
  \centering
    \includegraphics[width=6.9cm]{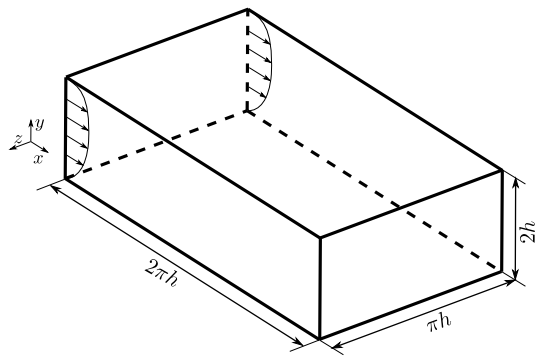}
  \caption{Sketch of the computational domain for the problem of modelling the
  turbulent flow in a plain channel.}
  \label{fig:channel_geometry}
\end{figure}

\textcolor{black}{The efficiency of the formulated computational procedure and
correctness of the proposed theoretical estimates are validated with two test
cases including the modelling of incompressible turbulent flow in a plain
channel~\cite{ref:Nikitin2006} and in a channel with a matrix of wall-mounted
cubes~\cite{ref:Meinders1998PhD, ref:Meinders1999, ref:VeldeERCOFTAC1999,
ref:MatheyERCOFTAC1999}. For the first case, the flow is modelled in a
rectangular computational domain of size $2\pi h \times 2h \times \pi h$
(Figure~\ref{fig:channel_geometry}), where $h$ is the channel half-height, with
periodic boundary conditions on streamwise and spanwise directions and no-slip
conditions on the channel walls. The constant flow rate is preserved during the
simulation with the Reynolds number $\mbox{Re}_b = 2800$, where $\mbox{Re}_b =
U_b h / \nu$ is defined using the bulk velocity $U_b$ and the channel
half-height $h$. The simulation is performed on the computational grid of $160
\times 140 \times 160$ cells (3.58~mln. cells). The grid is uniform in
streamwise and spanwise directions, and the grid stretching is applied in the
wall-normal direction with the ratio $\Delta y_{max} / \Delta y_{min} = 5$.
Introducing the viscous length scale $l_{\tau} = \nu / u_{\tau}$, where
$u_{\tau} = \sqrt{\tau_w / \rho}$ is the friction velocity, and $\tau_w$ is the
mean wall shear stress, one can express the grid spacings in wall units. The
corresponding values are: $\Delta x^+ = 6.95$, $\Delta y^+ = 0.81 - 4.1$,
$\Delta z^+ = 3.48$. The chosen grid resolution corresponds to the typical grid
spacings, used for DNS of turbulent channel flows, e.g.~\cite{ref:Nikitin2006,
ref:Nikitin2009}.}

\begin{figure}[htb]
  \centering
    \includegraphics[width=6.9cm]{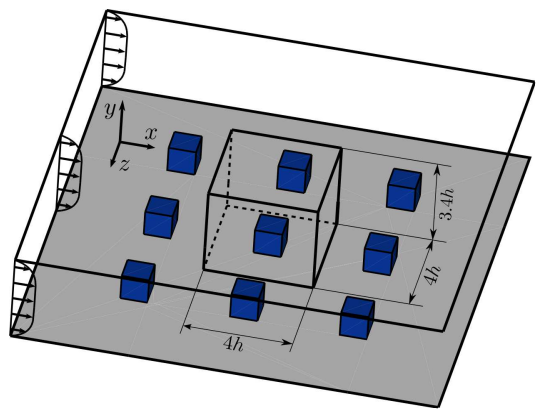}
  \caption{Sketch of the computational domain for the problem of modelling the
  turbulent flow in a channel with a matrix of wall-mounted cubes.}
  \label{fig:geometry}
\end{figure}

The second case considers the turbulent flow over the dedicated cube in the
channel (Figure~\ref{fig:geometry}), with periodic boundary conditions along the
two spatial directions and no-slip boundary conditions on the rigid walls.
The computational domain is set to $4h \times 3.4h \times 4h$, where $h$ is the
cube height. The constant flow rate is considered with the Reynolds number
$\mbox{Re}_b = 3854$, defined using the bulk velocity $U_b$ and the cube height
$h$. The computations are performed on three grids, consisting of approximately
2.32, 9.68, and 32.7~mln. cells. The parameters of the grids are specified
in~Table~\ref{tab:grid}. \textcolor{black}{The table also contains the $\Delta
y^+$ values for the grid spacings at the top wall of the channel, where the
viscous length scale is defined using the mean shear stress at this wall.
For the other walls the only local viscous length scales can be used to provide
reference information about the grid resolution. The mean grid spacings near the
walls are $\Delta x_i^+ \lesssim 2.4$, $1.34$, and $1.18$ for Grid~1, Grid~2,
and Grid~3 correspondingly. The maximum values are three times higher than the
mean ones, but they are observed in a few cells around the edges of the cube
only. The grids of 2.32 and 9.68~mln. cells are used for the detailed step by
step validation of the proposed computational methodology. For the finest grid
the only single run is performed to demonstrate the grid independence for the
obtained turbulent flow characteristics.}

The presented validation results were obtained on \textcolor{black}{two
supercomputers ``Lomonosov'' and ``Lomonosov-2''}. \textcolor{black}{The
``Lomonosov'' supercomputer contains the nodes with two 4-core Intel Xeon X5570
processors and 12~GB RAM, connected by QDR InfiniBand network.} The
``Lomonosov-2'' supercomputer is equipped with the nodes containing single
14-core Intel Xeon E5-2697v3 processor, \textcolor{black}{ 64~GB RAM,} and FDR
InfiniBand interconnect. Accordingly, the SparseLinSol library was configured to
perform the calculations \textcolor{black}{with 3-level hybrid model
(computational node / socket / core) for ``Lomonosov'' supercomputer and}
2-level hybrid model (computational node / core) for ``Lomonosov-2''
supercomputer, which allowed to utilize efficiently all available cores per
node. \textcolor{black}{The results presented below were obtained running 8 and
14~computational processes per node correspondingly, i.e. using each CPU core.}

\textcolor{black}{The effectiveness of the parallel computations is analyzed in
terms of parallel efficiency. This metric is defined as:
\begin{equation}
E(p) = \frac{T_1}{p \, T_p},
\end{equation}
where $p$ is the number of computing units, $T_1$ is the execution time using
single computing unit, and $T_p$ is the execution time using $p$ computing
units. The computing unit relates to the minimal number of computational
resources used to perform the simulation. For the second test case performed on
the coarse grid the computing unit is equal to single node and for the other
problems it is equal to 8~nodes.}


\subsection{Solution of SLAE with multiple RHS vectors}

The efficiency of the developed SLAE solver with multiple RHS vectors is
investigated for \textcolor{black}{three} matrices, which correspond to the
discretized pressure Poisson equation \textcolor{black}{for different test cases
and computational grids}. The BiCGStab iterative method with classical
algebraic multigrid preconditioner is used to solve the test systems. The
\textsc{PT-Scotch} graph partitioning library~\cite{ref:Chevalier2008} is
applied to build the sub-optimal matrix decomposition. In order to avoid the
slight variance in the number of iterations for convergence, the fixed number
of iterations is simulated for benchmarking purposes.

\textcolor{black}{The computational times for one iteration of SLAE solver and
the corresponding performance gains due to the use of operations with blocks of
vectors are presented in} Tables~\ref{tab:perf_gain_SLAE-case1}
and~\ref{tab:perf_gain_SLAE-case2}. As expected, the solution of SLAE with
multiple RHS vectors allows to improve the efficiency of the calculations and
reduce the computational time. The measured performance gains for both test
cases are in a good agreement with the proposed theoretical
estimate~\eqref{eq:mem_GSpMV_compact}, based on the memory traffic reduction.

\begin{table*}[htbp]
 \caption{\textcolor{black}{Theoretical and measured performance gains for
 solving SLAE with multiple RHS vectors. Test matrix for the problem of
 modelling turbulent flow in a plain channel, ``Lomonosov'' supercomputer,
 8~nodes. \label{tab:perf_gain_SLAE-case1}}}
 \centering
\textcolor{black}{
 \begin{tabular}{ | c | c | c | c |}
 \hline
  RHS & Time, s & Gain & Estimate~\eqref{eq:mem_GSpMV_compact} \\
  \hline
	1  & 0.05 & -  & -    \\
  \hline                                     
	2  & 0.07 & 1.45 & 1.48 \\
  \hline                                     
	4  & 0.10 & 1.86 & 1.95 \\
  \hline                                     
	8  & 0.18 & 2.12 & 2.21 \\
  \hline                                     
	16 & 0.32 & 2.45 & 2.29 \\
  \hline
\end{tabular}
}
\end{table*}

\begin{table*}[htbp]
 \caption{Theoretical and measured performance gains for solving SLAE with
 multiple RHS vectors. Test matrices for the problem of modelling turbulent flow
 in a channel with a matrix of wall-mounted cubes, ``Lomonosov-2''
 supercomputer. \label{tab:perf_gain_SLAE-case2}} 
 \centering
 \begin{tabular}{ | c | c | c | c | c | c |}
 \hline
  \multirow{2}{*}{RHS} & \multicolumn{2}{|c|}{Grid 1 (1 node)} &
  \multicolumn{2}{|c|}{Grid 2 (8 nodes)} &
  \multirow{2}{*}{Estimate~\eqref{eq:mem_GSpMV_compact}} \\
  \cline{2-5}
                & Time, s & Gain & Time, s & Gain & \\
  \hline
	1  & 0.12 & -    & 0.51 & -    & -    \\
  \hline                                     
	2  & 0.16 & 1.45 & 0.69 & 1.48 & 1.48 \\
  \hline                                     
	4  & 0.25 & 1.86 & 1.05 & 1.92 & 1.95 \\
  \hline                                     
	8  & 0.44 & 2.13 & 1.87 & 2.17 & 2.21 \\
  \hline                                     
	16 & 0.79 & 2.37 & 3.58 & 2.26 & 2.29 \\
  \hline
\end{tabular}
\end{table*}

\begin{figure*}[htbp]
  \centering
    \includegraphics[width=8.0cm]{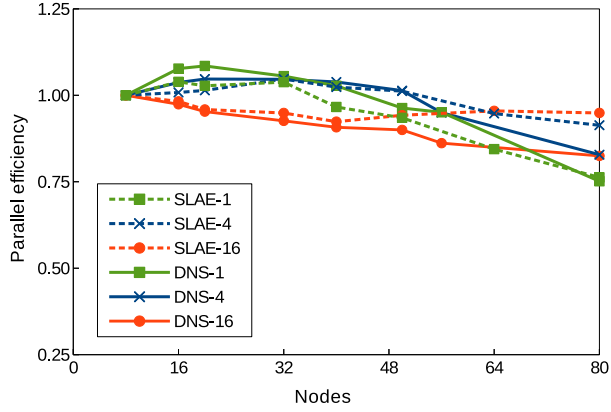}
  \caption{Parallel efficiency of the linear solver (SLAE) and the DNS
  application (DNS) performed with 1, 4, and 16~flow states. Test problem of
  modelling turbulent flow in a   plain channel, ``Lomonosov'' supercomputer.}
  \label{fig:par_efficiency_channel}
\end{figure*}

\begin{figure*}[htbp]
  \centering
    \includegraphics[width=8.0cm]{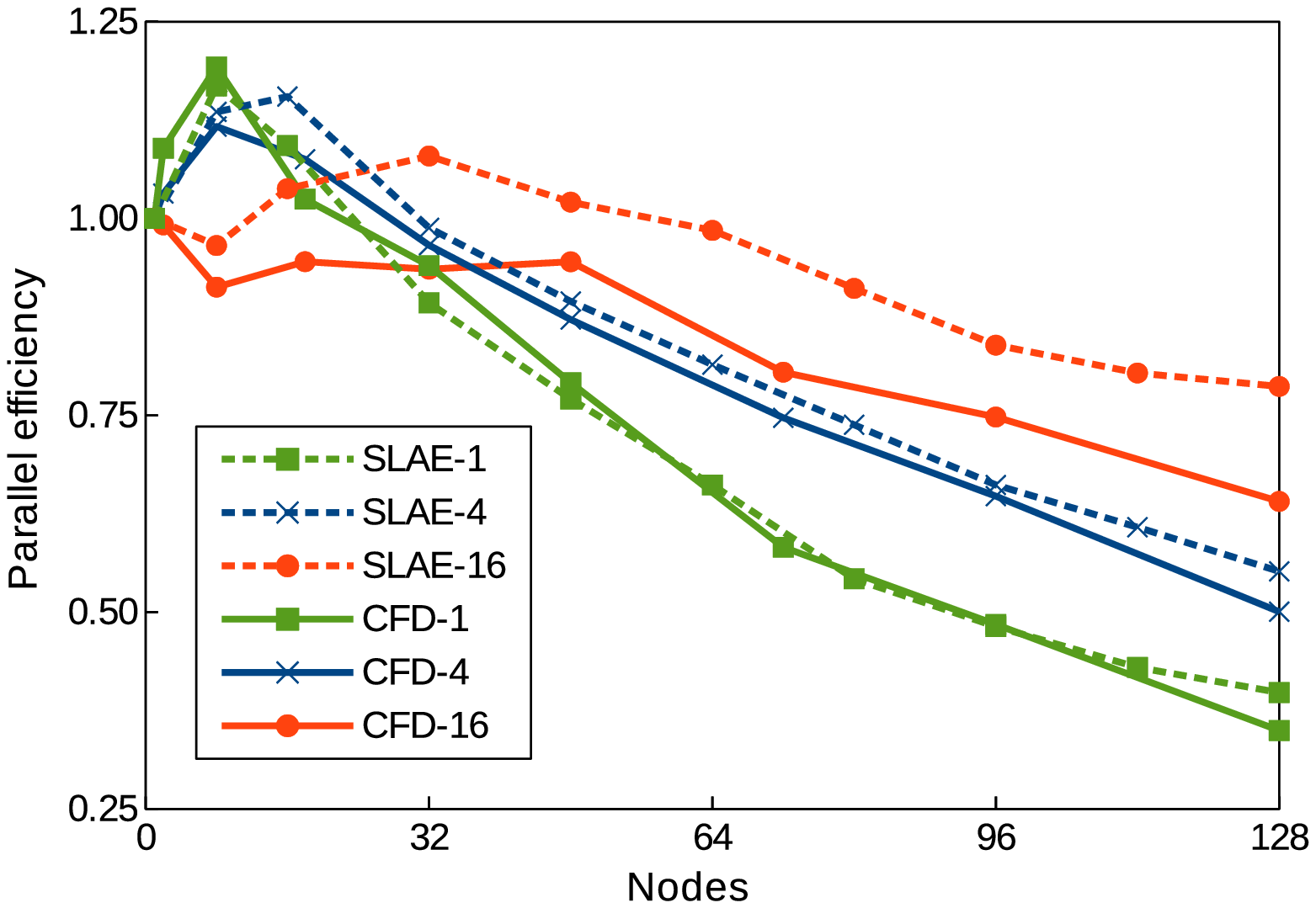}
    \includegraphics[width=8.0cm]{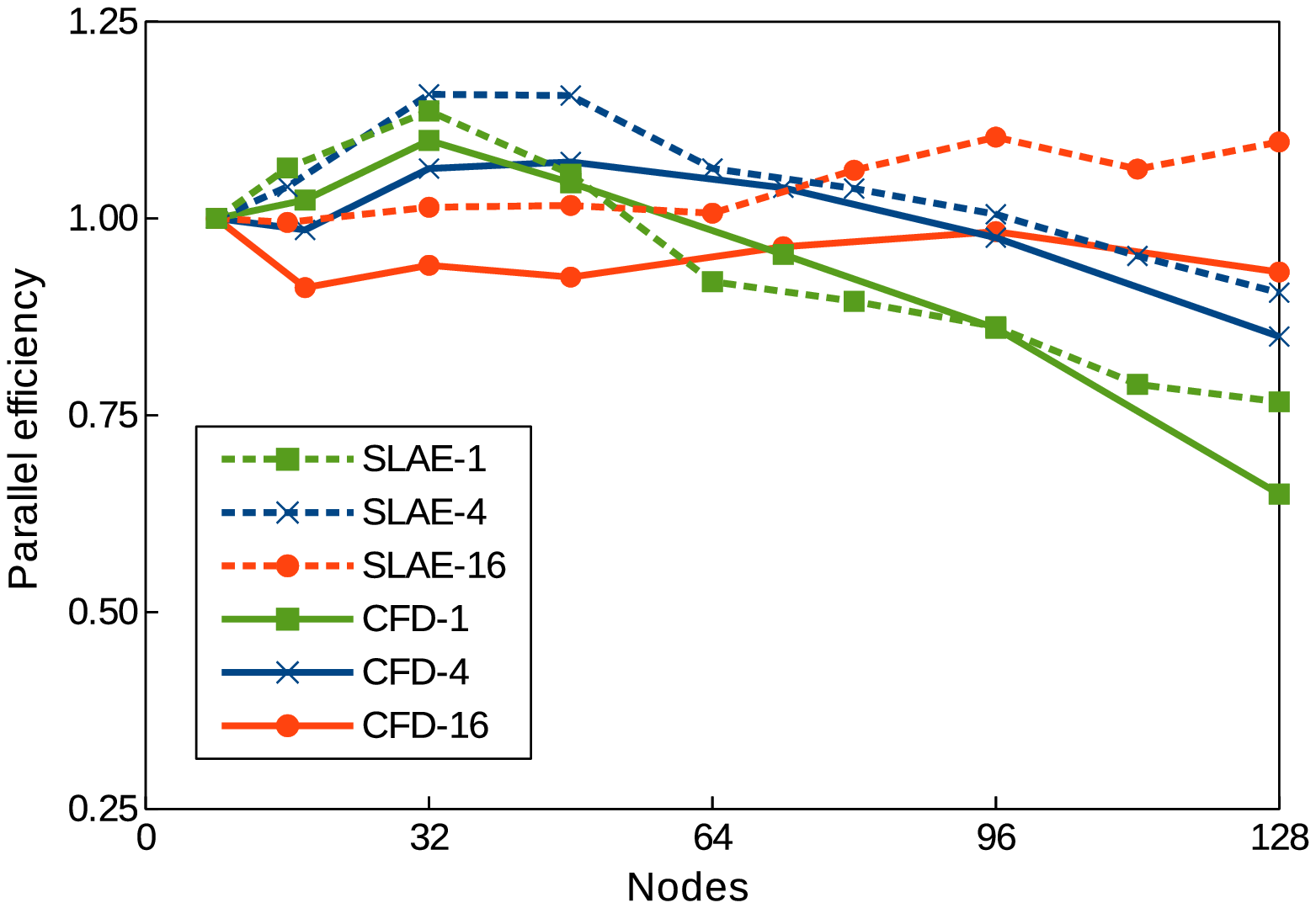}
  \caption{Parallel efficiency of the linear solver (SLAE) and the DNS
  application (DNS) performed with 1, 4, and  16~flow states. Test problem of
  modelling turbulent flow in a channel with a matrix of wall-mounted cubes,
  Grid~1~(left) and Grid~2~(right), ``Lomonosov-2'' supercomputer.}
  \label{fig:par_efficiency_cube}
\end{figure*}

The parallel efficiency results for 1, 4, and 16~RHS vectors for the cases
considered are presented in Figures~\ref{fig:par_efficiency_channel}
and~\ref{fig:par_efficiency_cube}. The plots demonstrate the improvement in
parallel efficiency for the simulations with multiple RHS vectors. The
scalability for single RHS vector degrade faster compared to other runs. For
example, the parallel efficiency of only 66\% for 64~nodes is obtained for the
test problem of 2.32~mln. cells, while for 4 and 16~RHS vectors it reaches about
81\% and 98\% respectively. This tendency affects the performance gain for
multi-node runs when solving SLAE with multiple RHS vectors: for 64~nodes it
increases up to 2.29 and 3.52 correspondingly, indicating an additional
advantage of using the operations with blocks of vectors for parallel runs.
The same behaviour is observed for two other test problems of 3.58 and 9.68~mln.
cells with only the difference that the degradation of parallel efficiency is
shifted towards the higher scales.

\subsection {Simultaneous modelling of multiple flow states}

\begin{table*}[htb]
 \caption{Theoretical and measured performance gains for modelling the
 single time step for multiple flow states. Test problem of modelling turbulent
 flow in a plain channel, ``Lomonsov'' supercomputer, 8 nodes.
 \label{tab:perf_gain_CFD_channel}}
 \centering
\textcolor{black}{
 \begin{tabular}{ | c | c | c | c |}
 \hline
  RHS & Time, s & Gain & Estimate~\eqref{eq:est2_1} \\
  \hline
	1  & 1.06 & -  & -    \\
  \hline                                     
	2  & 1.58 & 1.35 & 1.32 \\
  \hline                                     
	4  & 2.59 & 1.64 & 1.56 \\
  \hline                                     
	8  & 4.70 & 1.81 & 1.72 \\
  \hline                                     
	16 & 8.35 & 2.04 & 1.82 \\
  \hline
\end{tabular}
}
\end{table*}

\begin{table*}[htb]
 \caption{Theoretical and measured performance gains for modelling the time step
 for multiple flow states. Test problem of modelling turbulent flow in a channel
 with a matrix of wall-mounted cubes, ``Lomonosov-2'' supercomputer.
 \label{tab:perf_gain_CFD_cube}}
 \centering
 \begin{tabular}{ | c | c | c | c | c | c |}
 \hline
  \multirow{2}{*}{States} & \multicolumn{2}{|c|}{Grid 1 (1 node)} &
  \multicolumn{2}{|c|}{Grid 2 (8 nodes)} &
  \multirow{2}{*}{Estimate~\eqref{eq:est2_1}}
  \\
  \cline{2-5}
                & Time, s & Gain & Time, s & Gain & \\
  \hline
	1  & 2.78 & -    & 1.53  & -    & -    \\
  \hline                                     
	2  & 4.09 & 1.36 & 2.26  & 1.36 & 1.32 \\
  \hline                                     
	4  & 6.73 & 1.63 & 3.84  & 1.60 & 1.56 \\
  \hline                                     
	8  & 12.14 & 1.81 & 7.02 & 1.75 & 1.72 \\
  \hline                                     
	16 & 22.38 & 1.96 & 13.10 & 1.87 & 1.82 \\
  \hline
\end{tabular}
\end{table*}

\begin{table*}[t]
 \caption{\textcolor{black}{Memory consumption for simultaneous modelling of
 multiple flow states, in MB per computational core. Test problem of modelling
 turbulent flow in a channel with a matrix of wall-mounted cubes, Grid~1,
 ``Lomonosov-2'' supercomputer.}
 \label{tab:mem_CFD}}
 \centering
 \begin{tabular}{ | c | c | c | c | }
 \hline
  States & 1~node & 4~nodes & 16~nodes \\
  \hline
	1  & 330 & 145 & 162 \\
  \hline
	2  & 460 & 180 & 172 \\
  \hline
	4  & 720 & 250 & 190 \\
  \hline
	8  & 1170 & 385 & 230 \\
  \hline
	16 & 2120 & 655 & 312 \\
  \hline
  Approximation & $200 + 130 m$ & $110 + 35 m$ & $150 + 10 m$ \\
  \hline
\end{tabular}
\end{table*}

\begin{table*}[t]
 \caption{\textcolor{black}{Memory consumption for simultaneous modelling of
 multiple flow states, in MB per computational core. Test problem of modelling
 turbulent flow in a channel with a matrix of wall-mounted cubes, Grid~2,
 ``Lomonosov-2'' supercomputer.}
 \label{tab:mem_CFD2}}
 \centering
 \begin{tabular}{ | c | c | c | c | }
 \hline
  States & 8~nodes & 16~nodes & 32~nodes \\
  \hline
	1  & 305 & 222 & 276 \\
  \hline
	2  & 375 & 256 & 294 \\
  \hline
	4  & 505 & 324 & 330 \\
  \hline
	8  & 770 & 468 & 404 \\
  \hline
	16 & 1230 & 758 & 560 \\
  \hline
  Approximation & $240 + 65 m$ & $188 + 35 m$ & $258 + 18 m$ \\
  \hline
\end{tabular}
\end{table*}

A series of short runs performing several time steps for the DNS application is
calculated to investigate the simulation speedup when modelling multiple flow
states. As the input of the SLAE solver for pressure Poisson equation is
dominating in the overall simulation time, the strong correlation between the
parallel efficiency for the SLAE solver and the DNS application results is
expected. These expectations are confirmed by the corresponding calculations.

\begin{figure*}
  \centering
    \includegraphics[width=7.5cm]{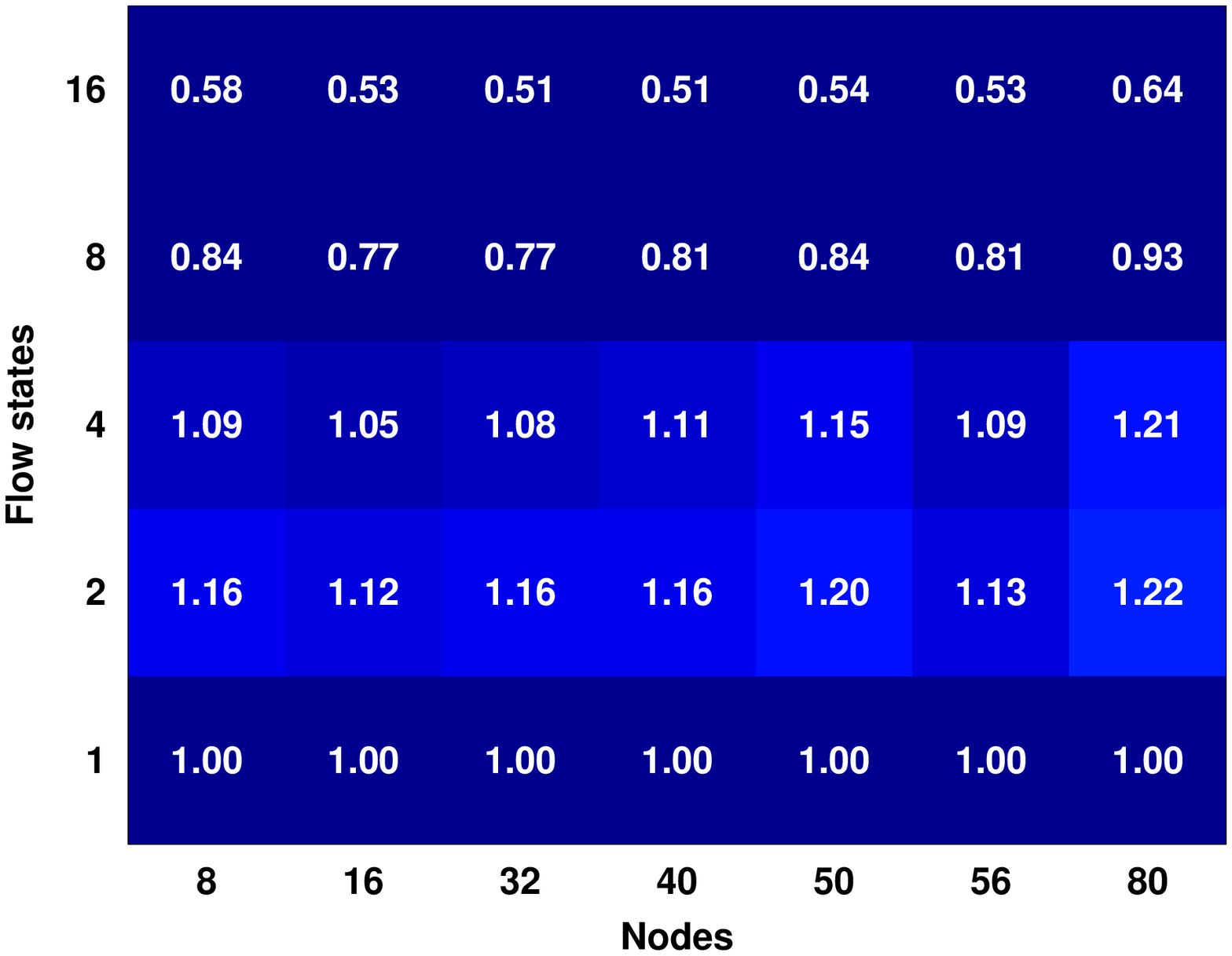}
    \includegraphics[width=7.5cm]{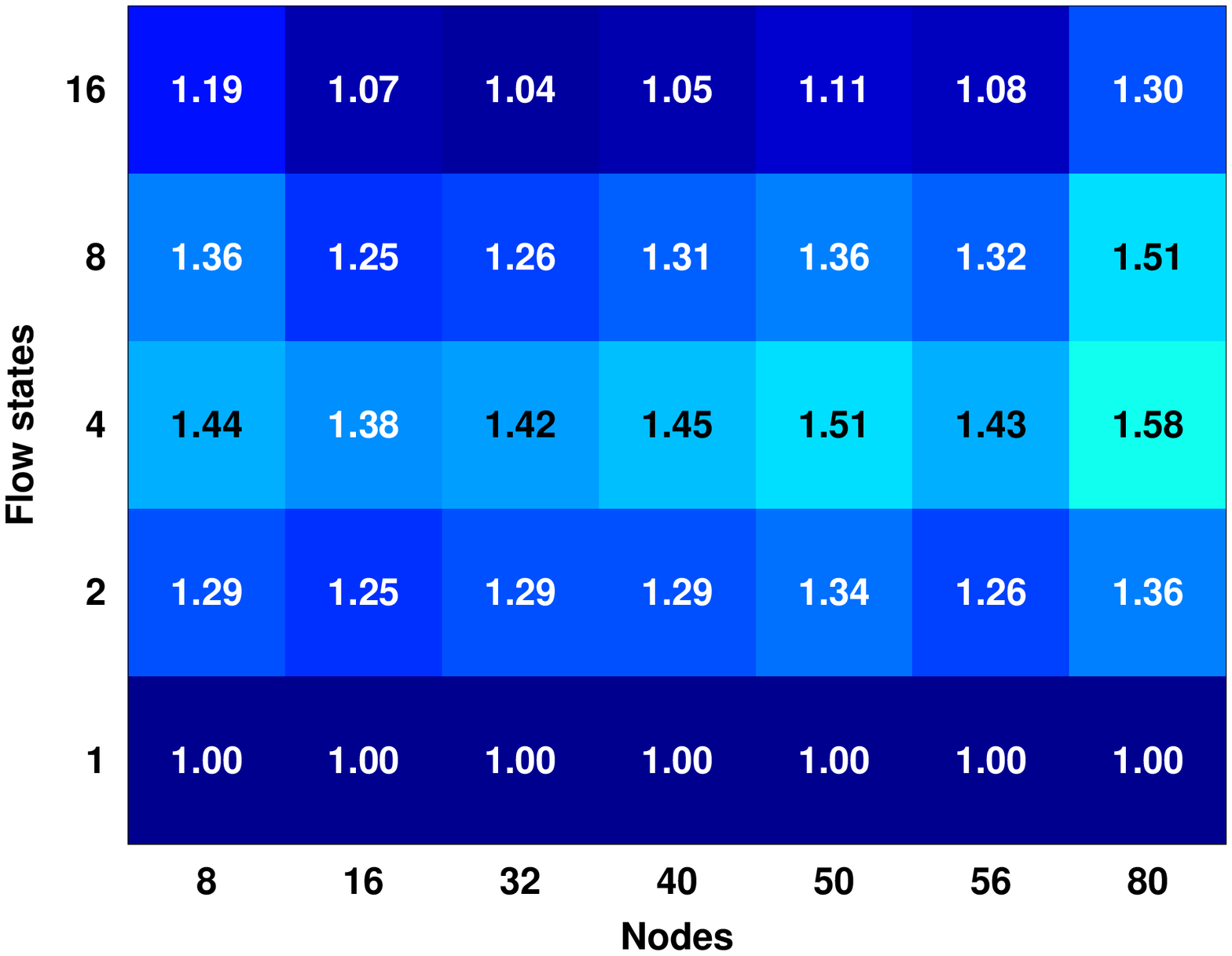}
  \caption{Overall performance gain estimate~\eqref{eq:est2} based on the single
  time step simulation results. Test problem of modelling turbulent flow in a
  plain channel. Left -- $\beta = 5$; right -- $\beta = 20$.}
  \label{fig:perf_gain_DNS_channel}
\end{figure*}

\begin{figure*}
  \centering
    \includegraphics[width=7.5cm]{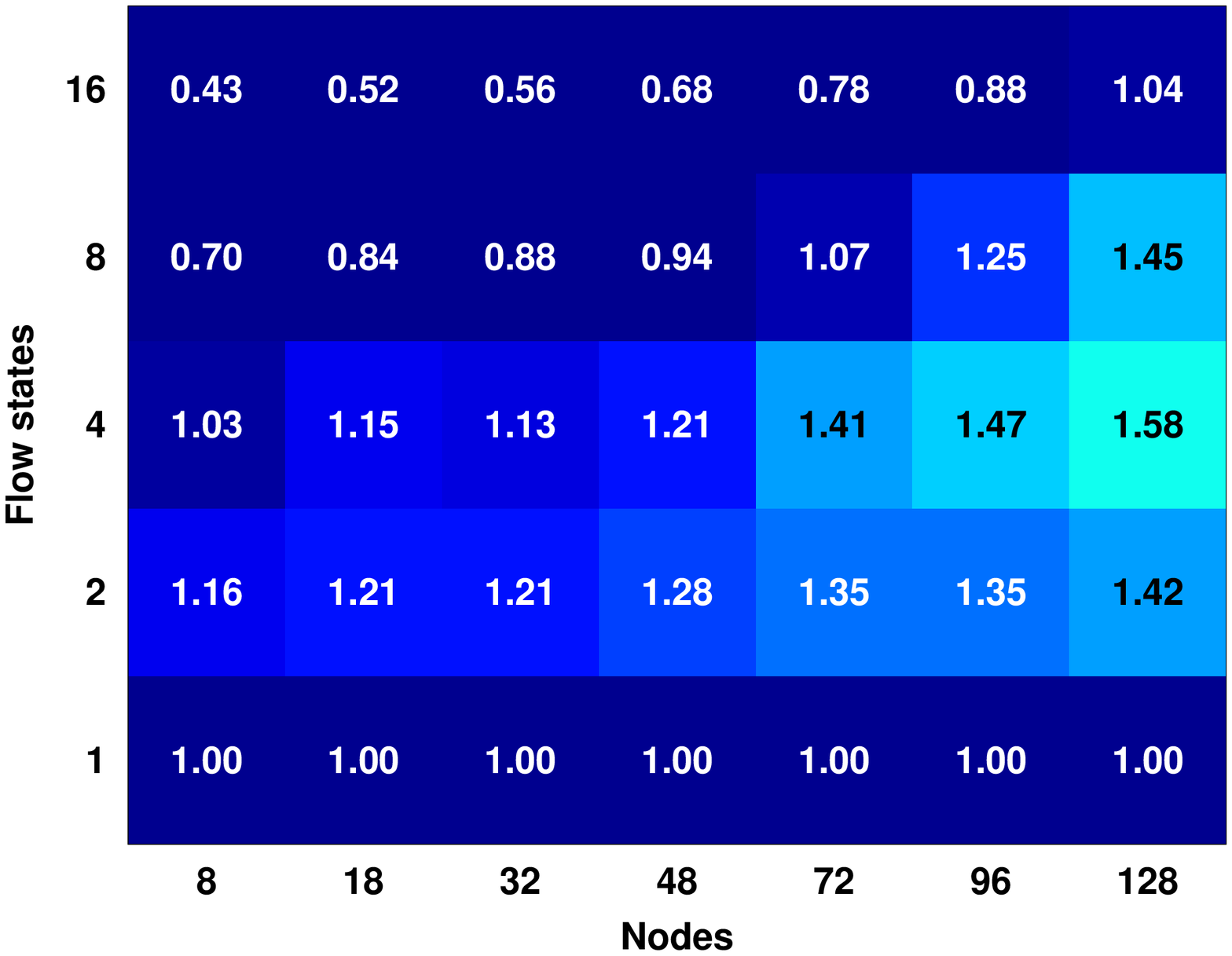}
    \includegraphics[width=7.5cm]{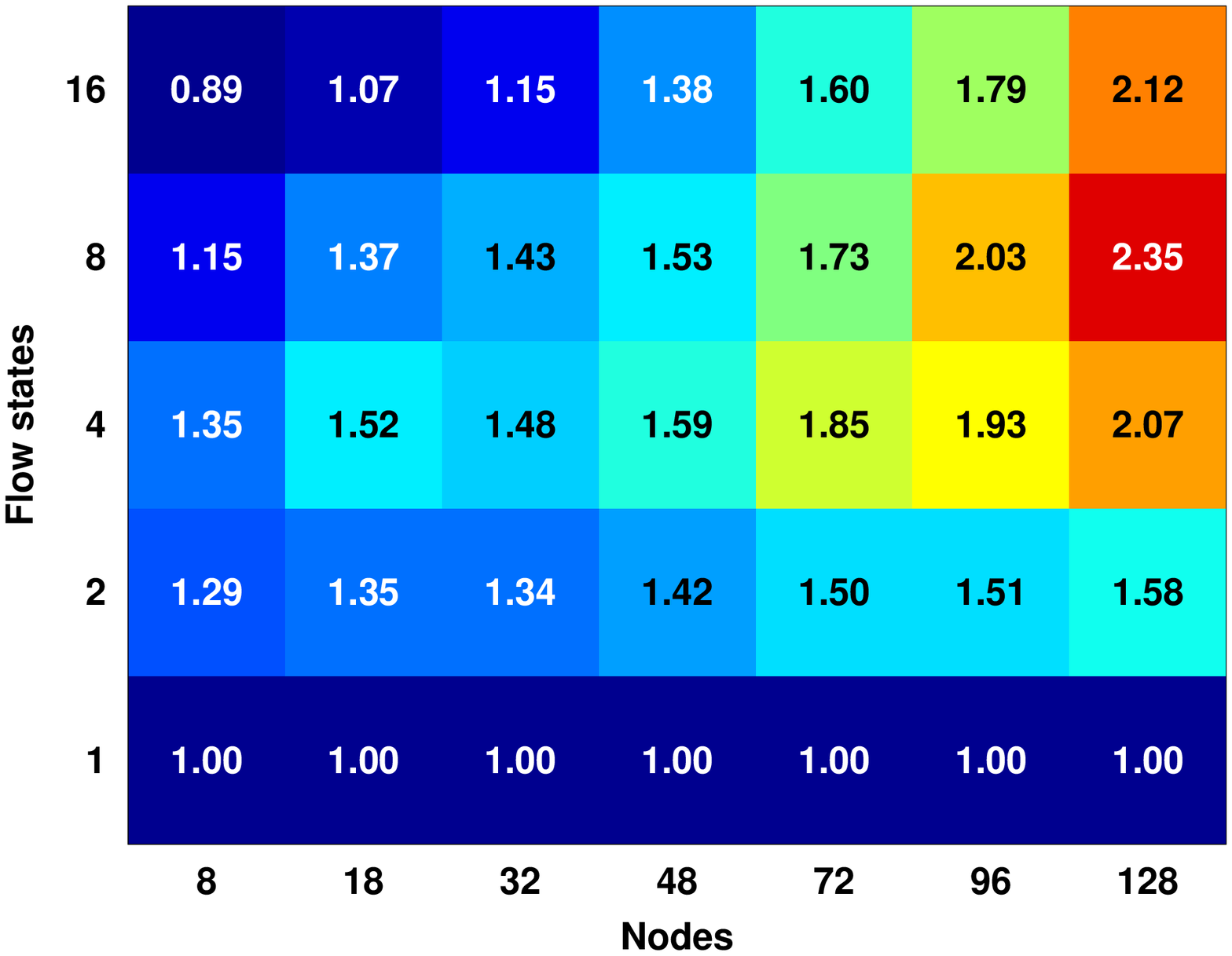}
    \includegraphics[width=7.5cm]{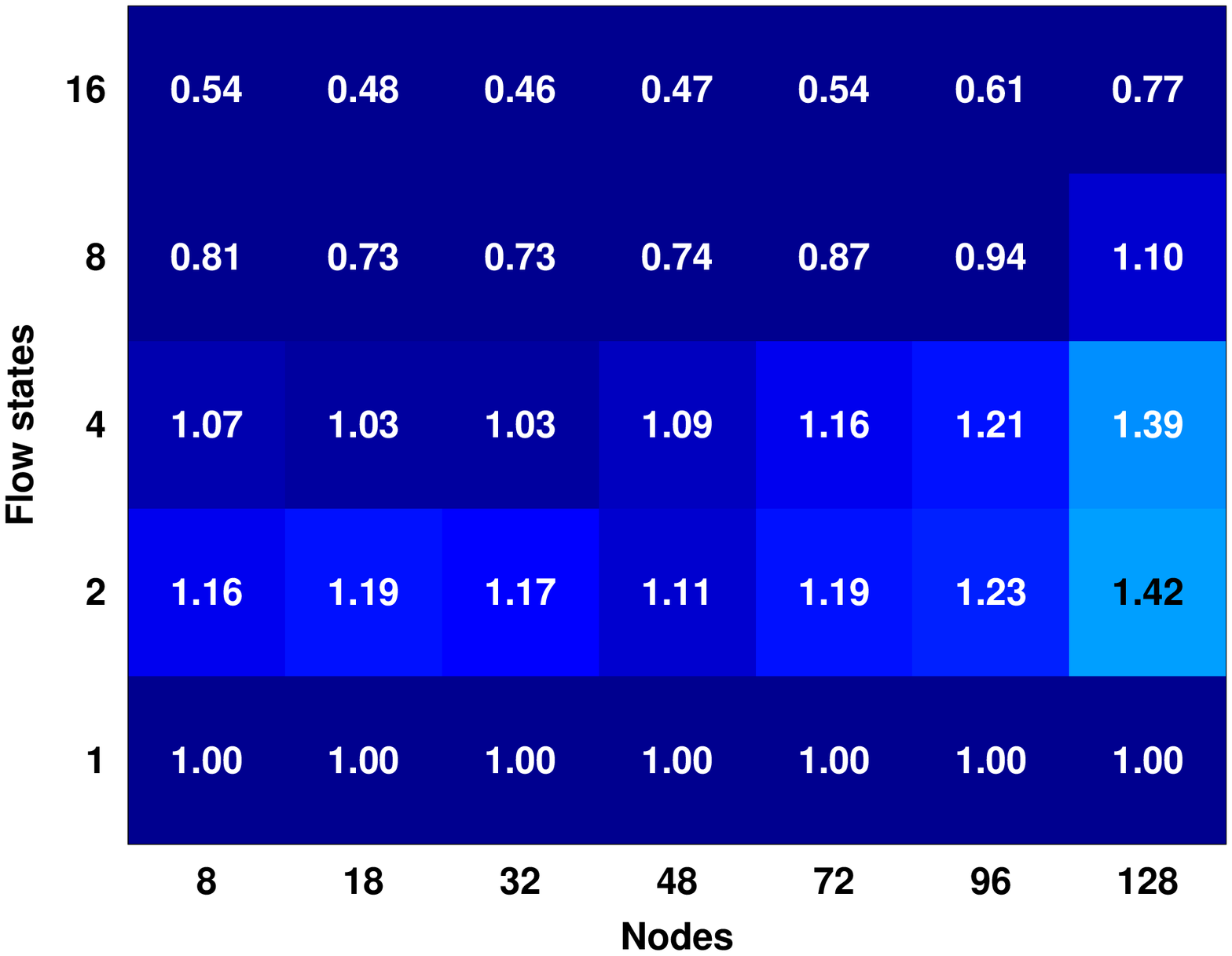}
    \includegraphics[width=7.5cm]{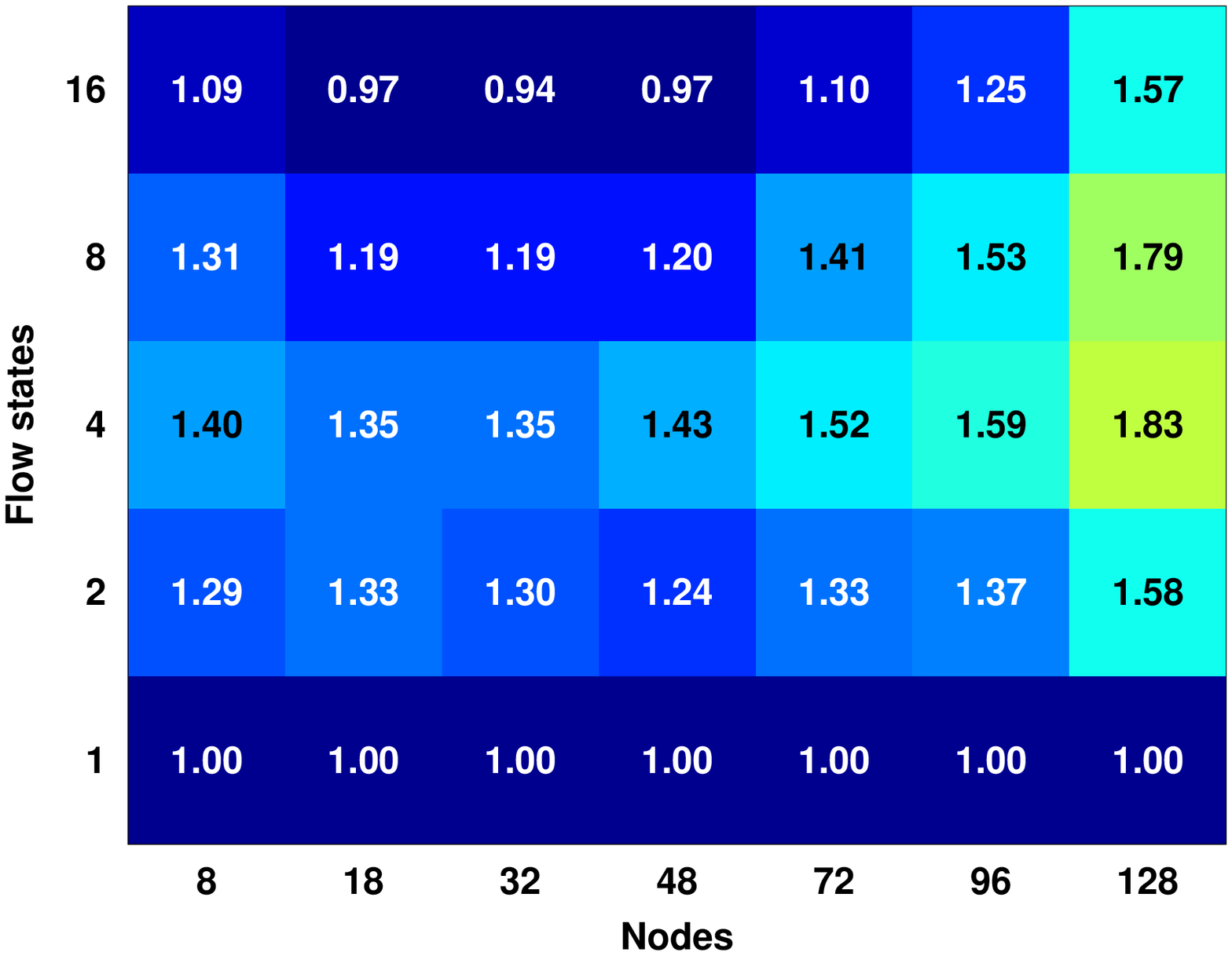}
  \caption{Overall performance gain estimate~\eqref{eq:est2} based on the single
  time step simulation results. Test problem of modelling turbulent flow in a
  channel with a matrix of a wall-mounted cubes. Top row -- results for Grid~1;
  bottom row -- results for Grid~2.
  Left column -- $\beta = 5$; right column -- $\beta = 20$.}
  \label{fig:perf_gain_DNS_cube}
\end{figure*}

The execution times per integration step, as well as performance gain results,
are summarized in Tables~\ref{tab:perf_gain_CFD_channel}
and~\ref{tab:perf_gain_CFD_cube}.
The obtained performance gains are similar for all the cases and computational
grids and correspond to the estimates~\eqref{eq:est2_1}. The parallel efficiency
results for the DNS application runs are presented in
Figures~\ref{fig:par_efficiency_channel} and~\ref{fig:par_efficiency_cube}.
They replicate the ones for the SLAE solver with only minor differences, mostly
caused by different data decomposition strategies used in standalone SLAE solver
and in DNS application.

\textcolor{black}{The simulations with multiple flow states increase the memory
consumption during the computations. The memory profiling is performed to
estimate the upper bound for the maximum number of simultaneously modelled flow
states in terms of memory consumption. The memory usage results for two
computational grids and with varying number of compute nodes are presented in
Tables~\ref{tab:mem_CFD} and~\ref{tab:mem_CFD2}. The obtained results can be
approximated by the simple formula:
\begin{gather}
M = \alpha + \gamma \cdot m,
\end{gather}
where $M$ is the memory consumption per core, $\alpha$ is the memory to store
the service information shared among all flow states (model description,
computational grid, SLAE solver data, etc.), and $\gamma$ is the memory to store
the data for the single flow state. The suggested approximation shows that the
second parameter, $\gamma$, is in proportion to the local grid block size and
decreases linearly with the number of nodes. The compute nodes of
``Lomonosov-2'' supercomputer provide about 4.6~GB of RAM per core. For example,
for the Grid~2 performed on 32~nodes (448~cores) the memory consumption per each
flow state does not exceed 18~MB/core (or 252~MB/node). This means the
simulations with several hundred flow states can be performed without in due
difficulties on ``Lomonosov-2'' compute nodes. The obtained profiling results
allow to state that the memory consumption does not provide any discernible
limitations on the applicability of the proposed approach.
}

The measured DNS application performance gain results are used to adjust the
overall simulation speedup estimates~\eqref{eq:est2}. The corresponding
distributions for the cases considered are presented in
Figures~\ref{fig:perf_gain_DNS_channel} and~\ref{fig:perf_gain_DNS_cube} for
$\beta=5$ and $\beta=20$. These results outperform the theoretical estimates in
Figure~\ref{fig:P_estimate} and show the potential of 2x~speedup for the overall
simulation.


\subsection {Full-scale DNS}

\textcolor{black}{
The several full-scale turbulent flow simulations for two test cases are
considered in this section to demonstrate the correctness of the proposed
theoretical estimates and real simulation speedup, and validate the simulation
results obtained as a result of averaging over different number of ensembles.
}

\subsubsection {DNS of turbulent flow in a plain channel}

\textcolor{black}{The two simulations are performed for the problem of
modelling turbulent flow in a plain channel. The overall simulation interval for
these runs is set to $T=3000$~time units, which comprises transition
interval of $T_T=500$~units and time averaging interval of $T_A=2500$~units with
the corresponding times ratio parameter $\beta = 5$. The initial velocity field
is specified in the form of laminar Poiseuille flow with periodic perturbations
for $x$- and $z$-velocity components along the spanwise and streamwise
directions:
\begin{gather}
u_i = \left(1 - \left( \frac{y-h}{h}\right)^2 \right) \left( \frac{3}{2} +
\delta_i \sin \left( \frac{2 \pi z}{L_z} \right) \right), \ i = 0,\ldots,m-1, \\
w_i = \delta_i \left(1 - \left( \frac{y-h}{h}\right)^2 \right) \sin \left(
\frac{2 \pi x}{L_x} \right), \ i = 0,\ldots,m-1,
\end{gather}
where $L_x = 2 \pi h$, $L_z = \pi h$, $\delta_i \in \left[ -0.1, 0.1 \right]$ is
a random number, and $i$ is the flow state index.
}

\textcolor{black}{The full-scale runs are simulated using 40~compute nodes with
single and two simultaneously modelled flow states. According to the estimates
presented in Figure~\ref{fig:perf_gain_DNS_channel}, performance gain by a
factor of 1.16 is expected for this test case configuration. The real simulation
times for these two runs are 480~min. and 412~min. correspondingly, that is the
performance gain is equal to~1.17. The estimate based on the single time step
simulation results correctly predicts the overall simulation speedup.
}

\begin{figure}
  \centering
    \includegraphics[width=7.8cm]{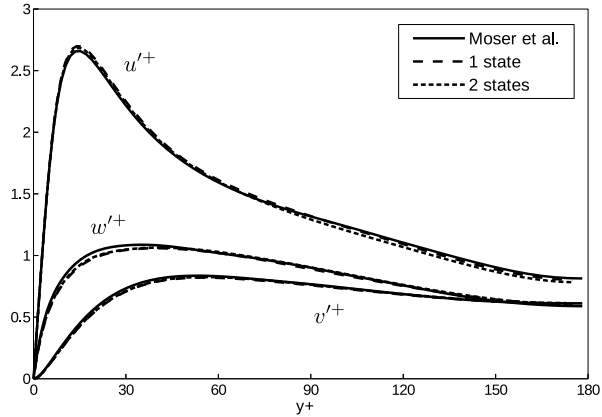}
  \caption{Rms velocity distributions for the simulations with averaging
  over single and two flow states, and the reference data of Moser et
  al.~\cite{ref:Moser1999}. Test problem of modelling turbulent flow in a plain
  channel.}
  \label{fig:turb_profiles_channel}
\end{figure}

\textcolor{black}{The obtained turbulent flow characteristics are compared with
the data, presented in~\cite{ref:Moser1999}. The friction Reynolds numbers,
$\mbox{Re}_{\tau} = u_{\tau} h / \nu$, for the performed simulations are
$\mbox{Re}_{\tau}^1 = 176.93$ and $\mbox{Re}_{\tau}^2 = 177.01$, which are
within 1\% error with respect to the reference value $\mbox{Re}_{\tau} =
178.13$. The comparison of the second order statistics includes distributions of
rms velocity profiles for three components obtained in two simulations together
with the data of Moser et al.~\cite{ref:Moser1999}. The results shown in
Figure~\ref{fig:turb_profiles_channel} are in good agreement with each other and
with reference data.
}

\subsubsection {DNS of turbulent flow over a matrix of wall-mounted cubes}

The six full-scale runs are performed when modelling the turbulent flow over a
matrix of wall-mounted cubes. In total, the interval of $T = 2100$ time units
is performed, which includes $T_T = 100$ units of the initial transition and
$T_A=2000$ units of averaging to collect the turbulent statistics.
This problem setup corresponds to the times ratio parameter $\beta = 20$.
\textcolor{black}{The preliminary streamwise velocity component in the interior
of the computational domain is specified in the form:
\begin{equation}
u_i = 1 + \delta_i \left( 1 - 2 \{ i/2 \} \right) \sin \left( \frac{2 \pi
z}{L_z} \left( 1 + [i/2] \right) \right), \ i = 0,\ldots,m-1,
\end{equation}
where $L_z = 4h$, $\delta_i \in \left[-0.05, 0.05 \right]$ is a random number,
$i$ is the flow state index, and $\left[ \alpha \right]$ and $\{ \alpha \}$ are
the integer and fractional parts of $\alpha$ respectively. To obtain initial
flow states the preliminary velocity distributions are projected on the
divergence-free space and adjusted to preserve the predefined flow rate.}

The two runs with single and four simultaneously modelled flow states
for Grid~1 are calculated on 32~nodes, with corresponding simulation times of
936~min. and 640~min. The obtained simulation speedup by a factor 1.46 coincides
with the adjusted estimate in Figure~\ref{fig:perf_gain_DNS_cube}, where the
performance gain of~1.48 is predicted. The third run for Grid~1 with modelling
of eight flow states is performed on 96~nodes. According to the estimates in
Figure~\ref{fig:perf_gain_DNS_cube} and the parallel efficiency results in
Figure~\ref{fig:par_efficiency_cube}, the simulation speedup by a factor of 3 is
expected compared to the one with single flow state performed on 32~nodes. The
execution time for this run is equal to 320~min., i.e. the speedup by a
factor of 2.95 is obtained. These calculations also demonstrate good
correspondence between the estimates and full-scale simulations.

\begin{figure}
  \centering
    \includegraphics[width=10.0cm]{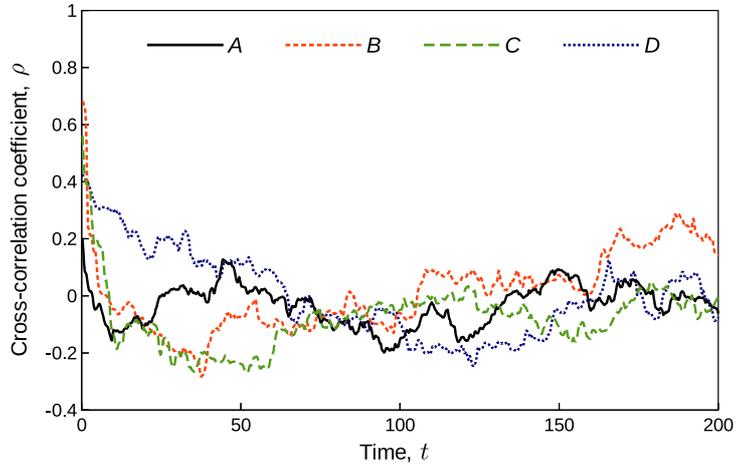}
  \caption{\textcolor{black}{Cross-correlation distributions for streamwise
  velocity evolution in 4~control points: $A=\{0.73h, \ h, \ 0\}$, $B=\{0.73h, \
  0.8h, \ 0\}$, $C=\{1.07h, \ 0.52h, \ 0\}$, $D=\{1.54h, \ 0.52h, \ 0.51h\}$;
  the origin of the coordinate system is located at the bottom wall of the
  channel in the middle of the cube. The time series length is equal to
  $T_0=50$.}}
  \label{fig:cube_2correlation}
\end{figure}

\begin{figure*}
  \centering
    \includegraphics[width=10cm]{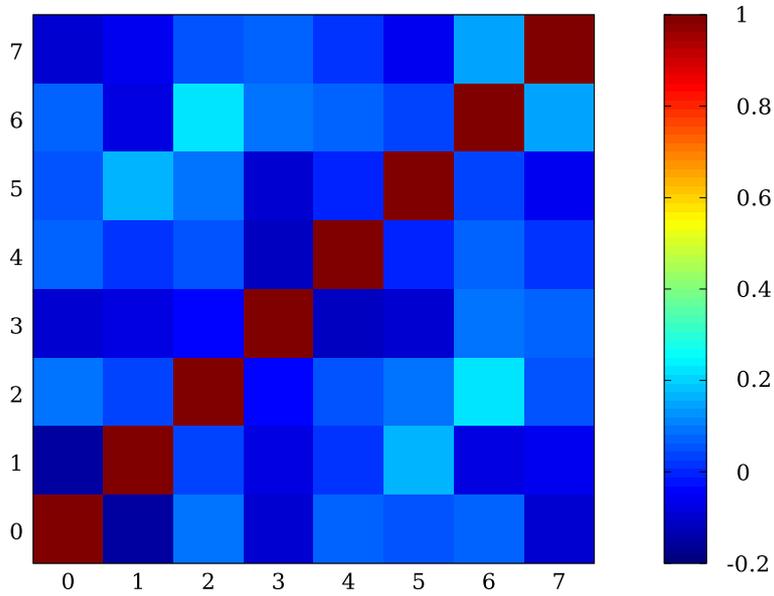}
  \caption{\textcolor{black}{Cross-correlation matrix for streamwise velocity for
  eight initial turbulent flow states ($t=T_T = 100$) for the data monitored in the
  control point $A=\{0.73h, \ h, \ 0\}$. Time series length is equal to $T_0 =
  50$.}} \label{fig:cube_8correlation}
\end{figure*}

\textcolor{black}{The cross-correlation analysis is performed for the third run
to demonstrate the correctness of the chosen transition interval and the fact
that the initial turbulent flow states are uncorrelated. The time series for the
velocity components monitored in several control points around and in the wake
of the cube are analyzed. The cross-correlation function is defined
as~\cite{ref:Chatfield2016}:
\begin{equation}
\rho_{XY} = \frac{cov \left( X,Y \right)}{\sqrt{var \left(X \right) var \left(
Y \right)}}, \label{eq:crosscorrelation}
\end{equation}
where $X$ and $Y$ are the segments of the velocities time series corresponding
to different flow states, $cov(X,Y)$ is the covariance of $X$ and $Y$, and
$var(X)$ is the variance of $X$. The time series $X$ and $Y$ are defined by two
parameters, $t$ and~$T_0$. The first one corresponds to the start time and the
second one corresponds to the length of the time series.
Figure~\ref{fig:cube_2correlation} shows the cross-correlation distributions for
the streamwise velocities obtained in four points for two flow states ($i=0$ and
$i=1$) and with $T_0=50$. The presented results vividly demonstrate that these
two flow states become uncorrelated at $t \sim 10$ which is lower than the
transition interval $T_T = 100$ chosen in the simulations. The cross-correlation
coefficients for eight flow states for $t = 100$ and $T_0=50$ are presented in
Figure~\ref{fig:cube_8correlation}. This distribution also demonstrates that all
eight initial turbulent flow states used for time averaging are uncorrelated.}

The two additional simulations are performed for the Grid~2 with 72~nodes for
single and four flow states. The measured execution times are 4011~min. and
2625~min. with corresponding speedup by a factor of~1.53. This value is also in
good agreement with the estimate, which predicted the performance gain equal
to~1.52.

\begin{figure*}
  \centering
    \includegraphics[width=7.5cm]{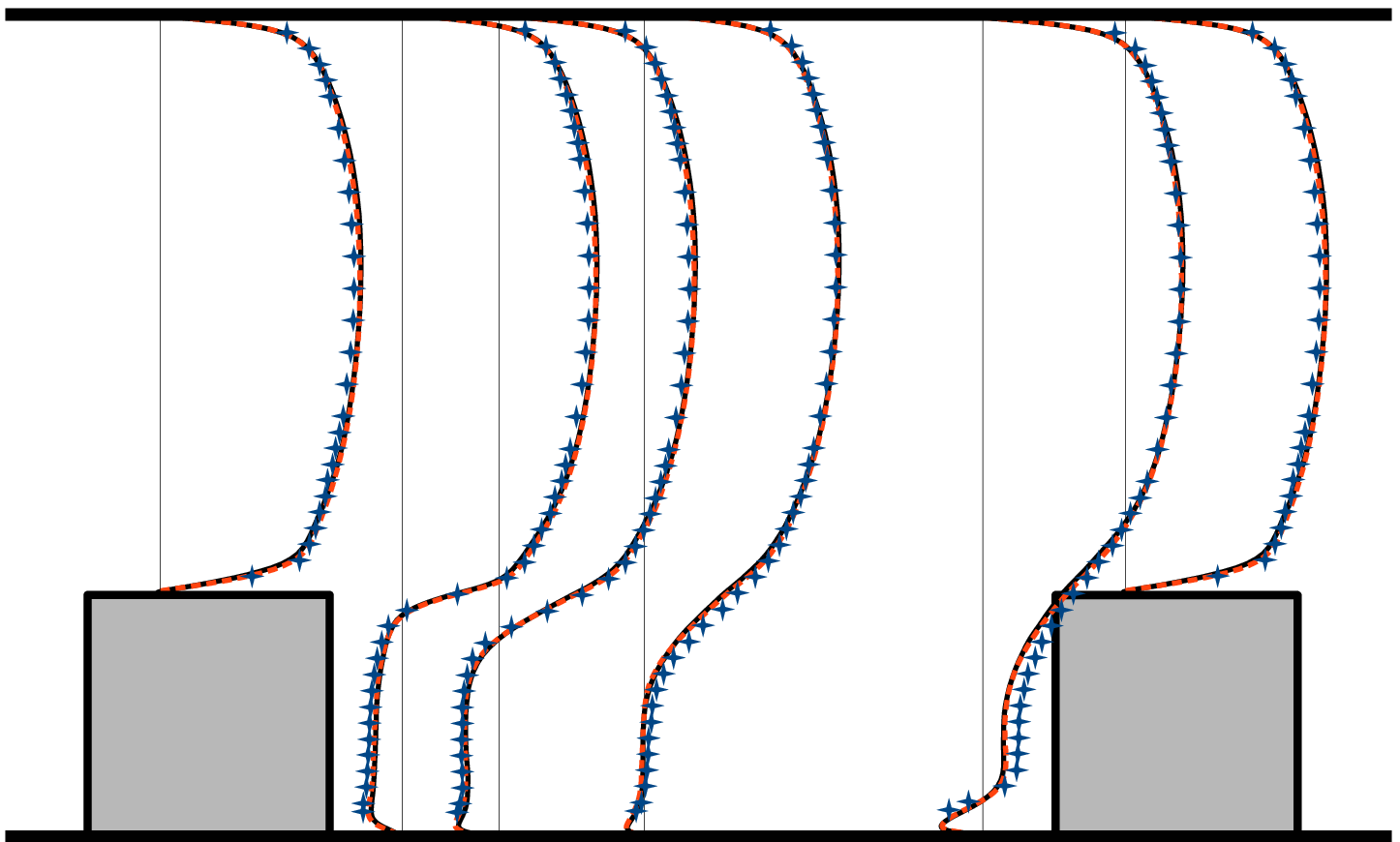}
    \includegraphics[width=7.5cm]{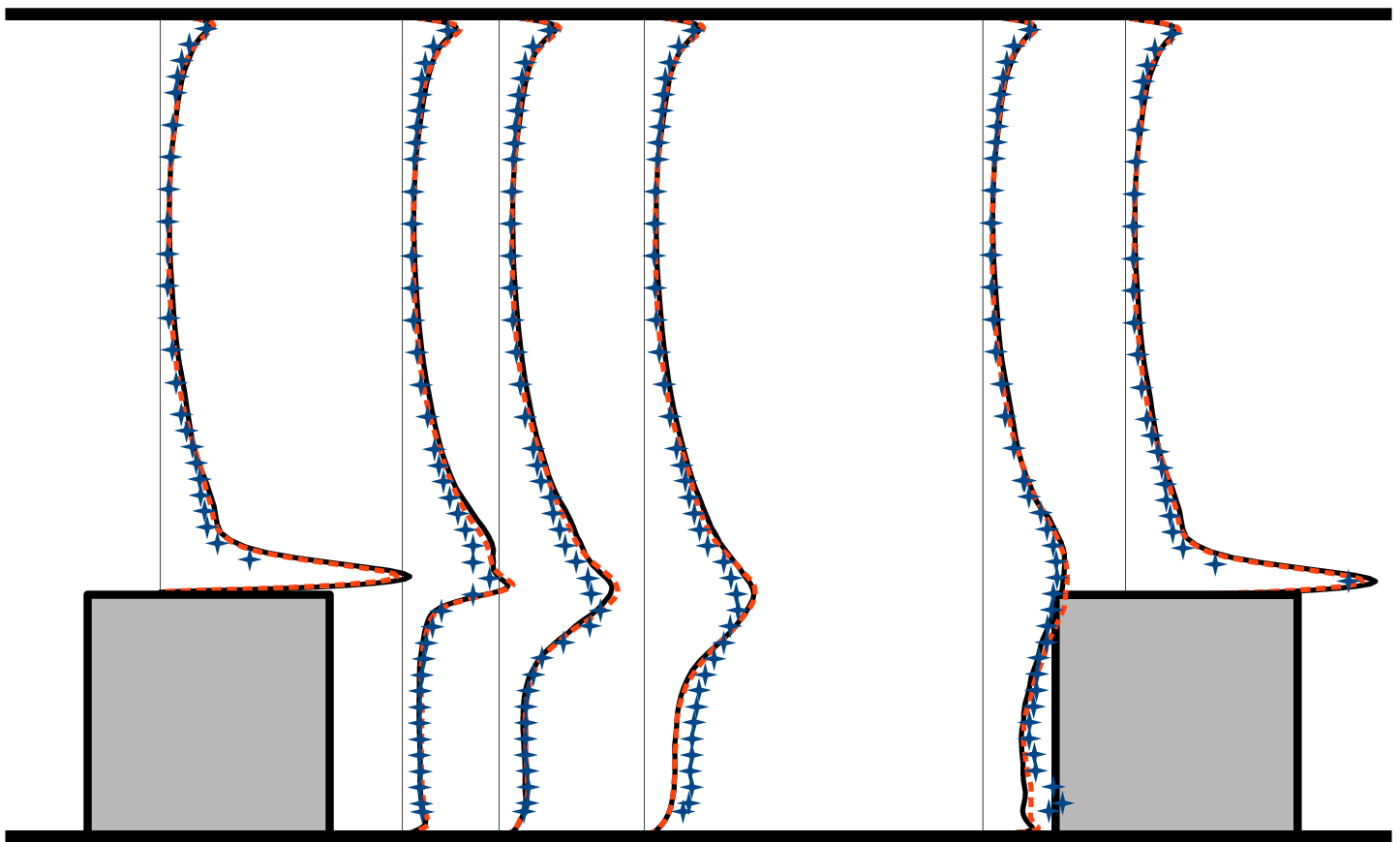}
  \caption{The mean streamwise velocity (left) and $\overline{u'^2}$ Reynolds
  normal stress (right) distributions at the channel cross section bisecting the
  cube. Continuous line -- Grid~1, averaging over single flow state; dashed line
  -- Grid~2, averaging over four flow states; markers -- experimental data.}
  \label{fig:turb_characteristics}
\end{figure*}

The turbulent flow characteristics, obtained in the simulations with
simultaneous modelling of multiple flow states, are compared with traditional
DNS results and with experimental data. The first and second order statistics in
the $x y$-plane bisecting the cube are inspected. The corresponding mean
streamwise velocity and Reynolds normal stress distributions along the channel for the
Grid~1 with single flow state, Grid~2 with four flow states and the ones
obtained in the experiments~\cite{ref:Meinders1998PhD, ref:Meinders1999} are
presented in Figure~\ref{fig:turb_characteristics}. The difference between the
simulation results for single and four flow states with different grids is
almost indistinguishable for both mean velocity and Reynolds normal stress
distributions. The observed variances between the numerical and experimental
data coincide with the ones for the numerical simulation results of other
authors~\cite{ref:VeldeERCOFTAC1999, ref:MatheyERCOFTAC1999}.

\begin{figure*}
  \centering
    \includegraphics[width=7.5cm]{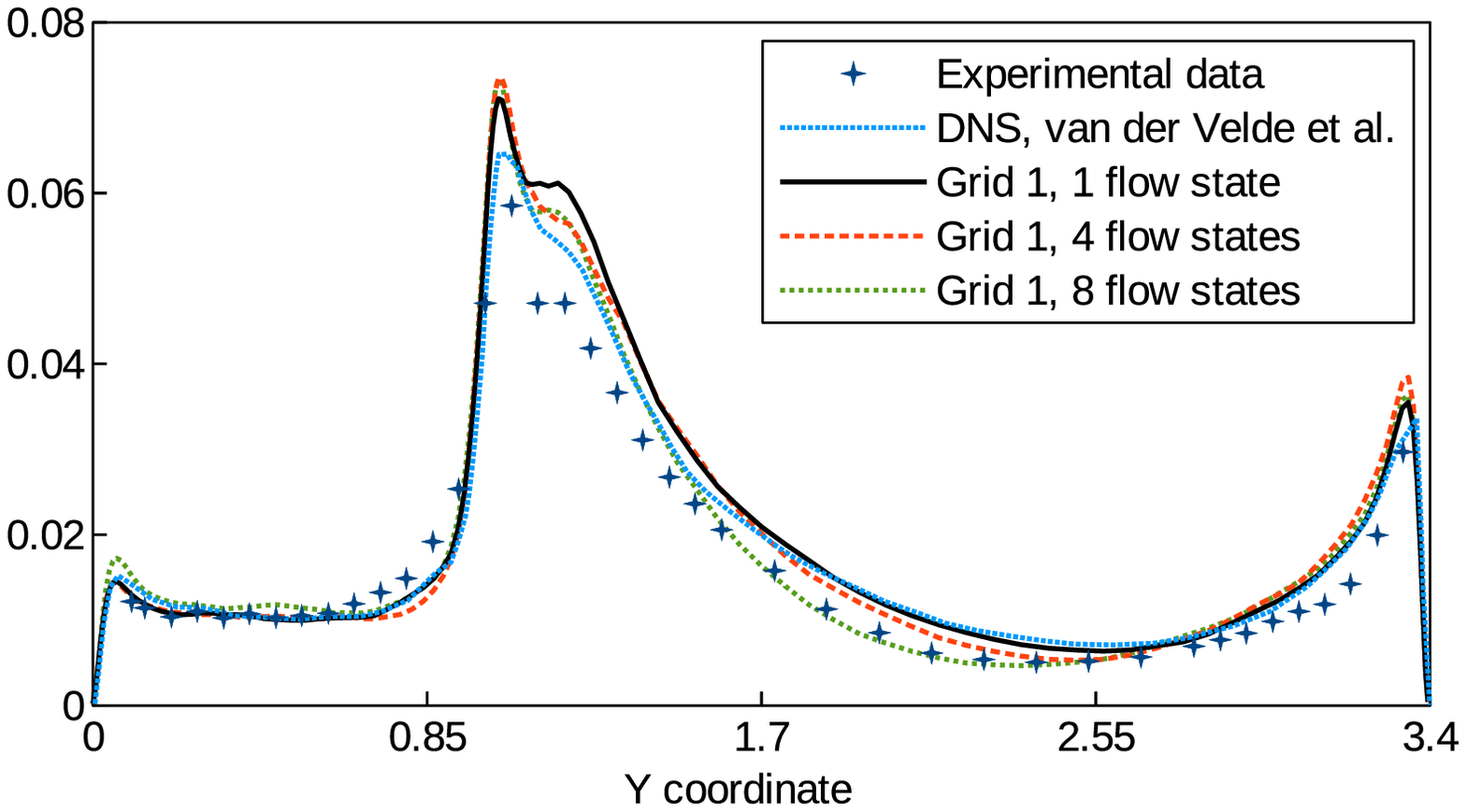}
    \includegraphics[width=7.5cm]{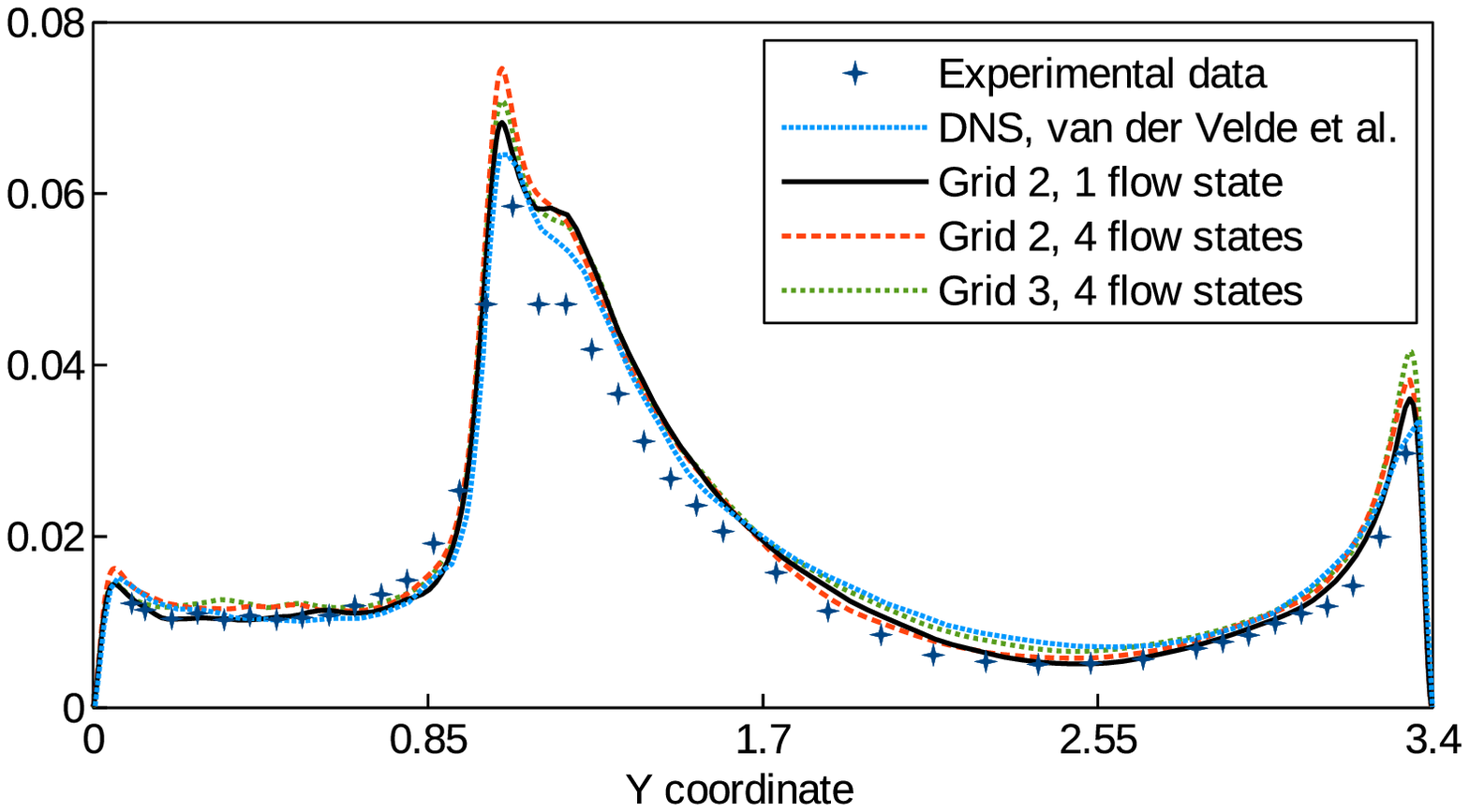}
  \caption{The $\overline{u'^2}$ Reynolds normal stress distributions at the
  distance $0.3h$ from the trailing edge of the cube. Comparison of numerical
  simulation results for various computational grids and number of flow states,
  experimental data of Meinders~\cite{ref:Meinders1998PhD}, and DNS results of
  van der Velde et al.~\cite{ref:VeldeERCOFTAC1999}.}
  \label{fig:turb_profiles_cube}
\end{figure*}

The streamwise Reynolds normal stress distributions for all the performed
simulations are compared in Figure~\ref{fig:turb_profiles_cube}. The
corresponding distributions located at the distance $0.3h$ from the trailing
edge of the cube are examined. \textcolor{black}{This figure also contains the
DNS results~\cite{ref:VeldeERCOFTAC1999} and experimental
data~\cite{ref:Meinders1998PhD}. The obtained simulation results are in good
correspondence with the DNS results of other authors. The slight variance in the
peak values of the Reynolds normal stress distributions for the numerical and
experimental data is observed. This fact could be a result of some experimental
measurements errors or minor differences in the experimental and numerical
problem statements.}

\begin{figure}[htbp]
  \centering 
  \includegraphics[width=7.8cm]{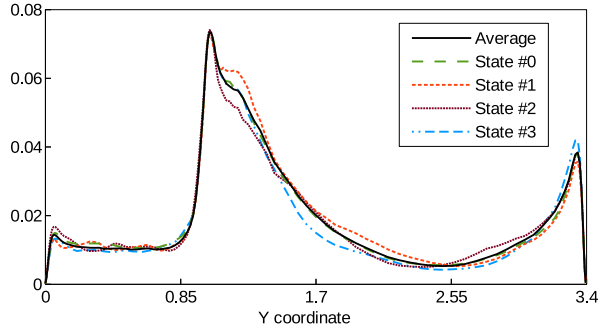}
  \caption{\textcolor{black}{The $\overline{u'^2}$ Reynolds normal stress
  distributions at the distance $0.3h$ from the trailing edge of the cube.
  Comparison of simulation results for four turbulent flow realizations
  performed in a single run, Grid~1.}}
  \label{fig:Urms_profile}
\end{figure}

\textcolor{black}{The simulations performed with varying number of flow states
demonstrate some deviations in the second order characteristics. The
corresponding distributions obtained as a result of averaging in time for each
of four flow realizations performed in a single run are presented in
Figure~\ref{fig:Urms_profile}. The figure demonstrates observable variance for
these distributions and suggests that better match between the single flow state
and multiple flow state simulations can be expected for a longer time averaging
interval.}

The presented results indicate that the proposed numerical procedure with
simultaneous modelling of multiple flow states allows to obtain a 1.5- to 2-fold
speedup for DNS applications compared to the conventional approach with averaging
over a single flow state due to better utilization of HPC hardware resources.


\section{Discussion}
\label{sec:discussion}

\textcolor{black}{The results presented in this paper can be considered as a
proof of concept for the simultaneous modelling of multiple turbulent flow
states approach to speedup the DNS/LES computations. Meanwhile, this paper not
focuses on several important aspects which may lead to further simulation
speedup improvements. These are briefly discussed in the current section.
}

\textcolor{black}{The presented results clearly demonstrate that the overall
simulation speedup is strongly dependent on the times ratio parameter, $\beta$.
An approach applied in this paper to obtain multiple independent initial
turbulent flow states produces significant overhead, thus reducing the benefits
of the proposed computational methodology. Various techniques can be applied
to reduce computational costs to obtain these initial flow states. Among them
can be an approach based on introduction of perturbations to the statistically
steady turbulent flow obtained for the single flow state. The simulations
on a series of nested computational grids with rescaling of results can also
help to provide multiple initial turbulent flow states with lower computational
costs.
}

\textcolor{black}{The validation results presented in this paper are obtained
on the supercomputers with Intel processors. The corresponding GSpMV operation
speedup coincides with the memory traffic reduction estimates and the expected
maximum is about 2.5. The results obtained in~\cite{ref:Imamura2016} show
the same values for the Intel-based compute system. However, for the system
with Sparc processors the observed performance gain is three times higher, about
7.6. This indicates the efficiency of the proposed method on some HPC systems
can significantly outperform presented in the paper performance gain results.
}

\textcolor{black}{An interesting observation is discussed
in~\cite{ref:Makarashvili2017}. The authors noted that the averaging over
multiple realizations of the same turbulent flow (ensemble averaging) may
provide higher accuracy compared to the conventional long-range simulation and
averaging in time. This results in a suggestion that the same accuracy of
statistics for multiple flow states simulations can be obtained by performing
shorter overall time averaging interval. This observation is not accounted in this
paper, but can be an additional source of improvement for the proposed
computational procedure.
}

\textcolor{black}{The current paper discusses the applicability of the
proposed approach to standard DNS or LES methods. These methods, however, do not
exhaust the range of possible applications to high-fidelity turbulent flow
modelling methods. For example, the quasi-DNS approach
(QDNS)~\cite{ref:Sandham2017} can be marked as a good candidate to use the same
methodology. The QDNS approach combines the main LES modelling with multiple
subscale near-wall DNS runs, performed to account the influence of the walls.
The corresponding DNS runs are independent and can be computed in a single run
with multiple flow states. Moreover, this algorithm has no drawbacks related to
simulation of transition interval to generate initial turbulent flow states for
these DNS runs.
}


\section{Conclusions}
\label{sec:conclusions}

The modified computational procedure for numerical simulation of turbulent flows
is presented. The suggested algorithm is based on the idea of simultaneous
modelling of multiple turbulent flow states followed by averaging of the
simulation results. The proposed approach allows to parallelize the transient
simulation in time and to use for the pressure Poisson equation the SLAE solver
operating with multiple right-hand side vectors. The simple theoretical
performance gain estimate, based on the memory traffic reduction for
matrix-vector operations with blocks of vectors, is formulated. The estimate
uses only two application-specific input parameters and allows to outline the
range of applicability for the proposed computational procedure and to predict
the simulation speedup. The speedup by a factor of 1.5 is expected with some
typical range of input parameters.

The extension of the SparseLinSol library, allowing to solve the systems of
linear algebraic equations with blocks of RHS vectors, and the application for
direct numerical simulation of turbulent flows are developed. The developed
software is applied to model the turbulent flow in the plain channel and in the
channel with a matrix of wall-mounted cubes, which are used to validate in
detail the formulated estimate and the proposed algorithm.

The step by step validation results of solving SLAE with multiple RHS vectors
and modelling the single DNS application time step show good correspondence
between the theoretical and numerical results. For several nodes runs the
speedup by a factor of 1.5 is observed. The larger scale runs including several
tens of compute nodes outperform the estimate thanks to the better
scalability of the SLAE solver operating with multiple RHS vectors.

Several full-scale DNS runs for two test cases, various computational grids,
variable number of compute nodes and computing systems, and number of modelled
flow states are performed. The observed performance gain results correspond to
the estimates, and the overall simulation speedup by a factor of 2 is
demonstrated. The presented turbulent flow characteristics for the full-scale
runs coincide with each other and are in agreement with the experimental data
and results presented by other authors. These facts vividly demonstrate the
correctness and the efficiency of the proposed numerical procedure for modelling
incompressible turbulent flows and the potential of 2x~speedup for large-scale
turbulent flow simulations.

\section*{Acknowledgments}
The author acknowledges Alexey Medvedev, Dr.~Alexander Lukyanov and
Dr.~Nikolay Nikitin for in-depth discussions of materials presented in this
paper. The research was carried out using the equipment of the shared
research facilities of HPC computing resources at Lomonosov Moscow State
University.

\bibliographystyle{elsarticle-num}

\bibliography{base}

\end{document}